%% file: main.tex
\documentclass[sigconf]{acmart}
\AtBeginDocument{%
  \providecommand\BibTeX{{%
    \normalfont B\kern-0.5em{\scshape i\kern-0.25em b}\kern-0.8em\TeX}}}

\setcopyright{acmcopyright}

\copyrightyear{2024}
\acmYear{2024}
\setcopyright{rightsretained}
\acmConference[CHI '24]{Proceedings of the CHI Conference on Human Factors in Computing Systems}{May 11--16, 2024}{Honolulu, HI, USA}
\acmBooktitle{Proceedings of the CHI Conference on Human Factors in Computing Systems (CHI '24), May 11--16, 2024, Honolulu, HI, USA}
\acmDOI{10.1145/3613904.3642068}
\acmISBN{979-8-4007-0330-0/24/05}

\acmSubmissionID{7980}

\usepackage{balance}       
\usepackage{graphics}      
\usepackage[T1]{fontenc}   
\usepackage{color}
\usepackage{booktabs}
\usepackage{textcomp}
\usepackage{eurosym}
\usepackage{ljhnick}
\usepackage{mathtools}
\usepackage{multicol}
\usepackage{tabularx}
\usepackage{makecell}
\usepackage{multirow}
\usepackage{array}
\usepackage{subcaption}
\usepackage{graphicx}
\usepackage{fontawesome5}
\usepackage{listings}
\usepackage{csquotes}
\usepackage{amsmath}
\usepackage{diagbox}
\usepackage{xcolor}
\usepackage{environ}
\usepackage{wrapfig}

\newenvironment{colortext}[1][black]{\color{#1}}{}

\NewEnviron{coloredtable}[1]{%
  \color{#1}%
  \BODY 
}

\newcolumntype{C}[1]{>{\centering\arraybackslash}p{#1}}

\newcolumntype{C}[1]{>{\centering\let\newline\\\arraybackslash\hspace{0pt}}m{#1}}

\begin{document}

\input{content/title}

\author{Jiahao Nick Li}
\affiliation{%
  \institution{Reality Labs Research, Meta \& UCLA}
  \city{Toronto}
  \country{Canada}}
\email{ljhnick@ucla.edu}
\orcid{0000-0002-4937-0024}

\author{Yan Xu}
\affiliation{%
  \institution{Reality Labs Research, Meta}
  \city{Redmond}
  \country{United States}}
\email{yanx@meta.com}

\author{Tovi Grossman}
\affiliation{%
  \institution{University of Toronto}
  \city{Toronto}
  \country{Canada}}
\email{tovi@dgp.toronto.edu}

\author{Stephanie Santosa}
\affiliation{%
  \institution{Reality Labs Research, Meta}
  \city{Toronto}
  \country{Canada}}
\email{ssantosa@meta.com}

\author{Michelle Li}
\affiliation{%
  \institution{Reality Labs Research, Meta}
  \city{Toronto}
  \country{Canada}}
\email{michelleli@meta.com}


\begin{abstract}

\input{content/00_abstract}
\end{abstract}


\input{content/00_teaser}



\maketitle

\input{content/01_intro}
\input{content/02_related}

\input{content/03_formative_study}

\input{content/04_diary_study}
\input{content/05_design_space}
\input{content/06_system}
\input{content/07_user_study}
\input{content/08_discussion}
\input{content/09_conclusion}


\bibliographystyle{ACM-Reference-Format}
\bibliography{reference}

\input{content/99_appendix}

\end{document}

%% file: content/title.tex




\title{\codename: Predicting Digital Actions in Response to Real-World Multimodal Sensory Inputs with LLMs}

%% file: content/00_abstract.tex
\begin{colortext}
The progression to ``Pervasive Augmented Reality'' envisions easy access to multimodal information continuously. 
However, in many everyday scenarios, users are occupied physically, cognitively or socially. This may increase the friction to act upon the multimodal information that users encounter in the world. 
To reduce such friction, future interactive interfaces should intelligently provide quick access to digital actions based on users' context.
To explore the range of possible digital actions, we conducted a diary study that required participants to capture and share the media that they intended to perform actions on (e.g., images or audio), along with their desired actions and other contextual information. 
Using this data, we generated a holistic design space of digital \textit{follow-up} actions that could be performed in response to different types of multimodal sensory inputs.
We then designed \codename, a pipeline powered by large language models (LLMs) that processes multimodal sensory inputs and predicts follow-up actions on the target information grounded in the derived design space. 
Using the empirical data collected in the diary study, we performed quantitative evaluations on three variations of LLM techniques (intent classification, in-context learning and finetuning) and identified the most effective technique for our task. 
Additionally, as an instantiation of the pipeline, we developed an interactive prototype and reported preliminary user feedback about how people perceive and react to the action predictions and its errors. 
\end{colortext}

%% file: content/00_teaser.tex
\begin{teaserfigure}
    \centering
    \includegraphics[width=1\linewidth]{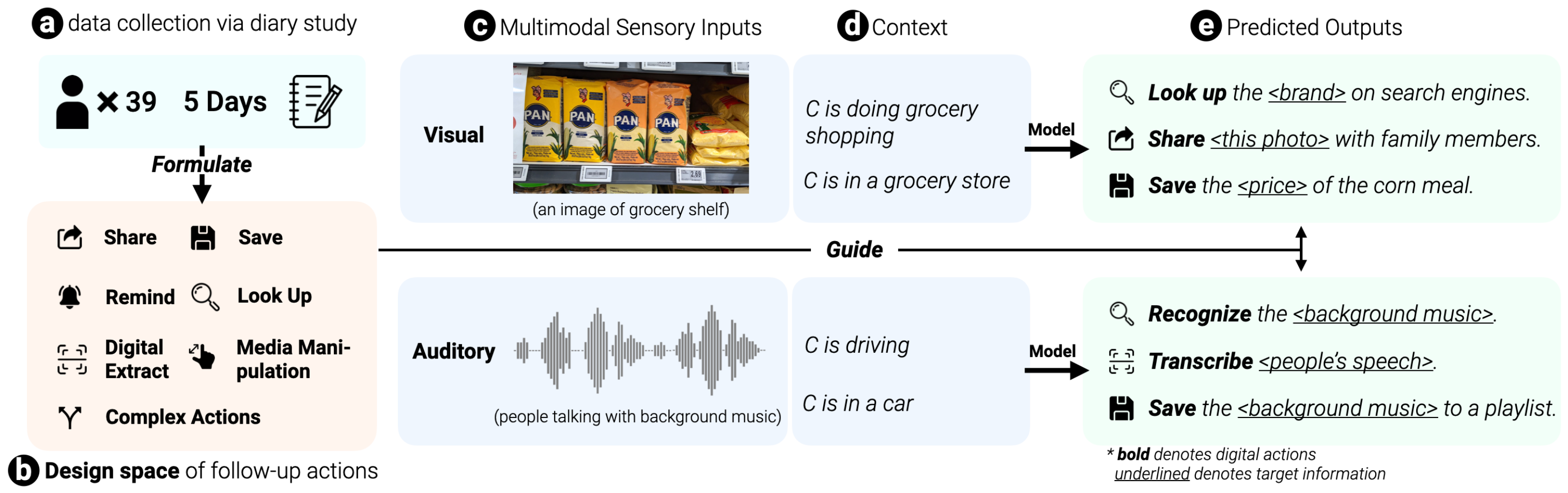}
    \caption{
    \textcolor{black}{\codename contributes: (1) a design space of digital follow-up actions (b) derived from data collected during a five-day diary study with 39 participants (a), and (2) a pipeline that takes multimodal sensory data (c) and contextual information (d) as inputs, and predicts what digital actions users might take and on which specific information in the input they might take these actions (e). The action prediction is guided by the design space.}
    }
    \label{fig:teaser}
\end{teaserfigure}

%% file: content/01_intro.tex
\section{Introduction}

\input{figures/flowchart}

\begin{colortext}

The progression towards ``Pervasive Augmented Reality (AR)'' envisions easy access to information in different modalities such as text, images, or audio, anytime and anywhere \cite{grubert2016towards}.
However, in many everyday scenarios within the real world, users are occupied physically, cognitively or socially, which may limit the use of typical AR inputs such as hand gestures and speech. 
This can present significant friction in interacting further with the information they encounter in the world.
For example, a driver noticing a movie billboard faces increased friction in (1) identifying the movie's name from the billboard and (2) searching for more details about the movie, due to the cognitive and physical demands of driving.
This motivates in the need for future interfaces to intelligently reduce friction in interacting with information \cite{ashbrook2010enabling}.

Interactions with real-world information generally involve two steps:
(1) retrieving desired information (\eg select the text on the billboard) and 
(2) performing corresponding \textit{follow-up actions} (\eg searching for more details on Google).
We envision that future interfaces should be designed to simultaneously process multimodal sensory inputs, analogous to \textit{human sensory perception}, and proactively suggest follow-up actions on the target information. 
This vision represents a more \textit{generalized} approach 
than 
\textcolor{black}{
existing approaches like iOS' text-in-a-photo action suggestions\footnote{https://support.apple.com/en-us/HT212630}, Google Lens\footnote{https://lens.google/}, or Shazam's song recognition\footnote{https://www.shazam.com/}, which recognize one specific modality of sensory inputs (\eg structured text, images, or audio) and map it to hard coded predefined actions (\eg detecting an address and launching a navigation app). 
}
However, to implement this more generalized vision, two main limitations need to be addressed:
\one existing systems cannot predict follow-up actions on aggregated data from multiple modalities and 
\two there is a limited understanding of the range of actions users intend to perform during real-world scenarios when using multiple modalities.
The latter is crucial for guiding the design of such systems, as it ensures that their output is grounded in a known action space, thus enabling the actions to be executable by the system.
Prior work has explored the design space of mobile and in-situ information needs \cite{church2014large,dearman2008examination}, \ie
\textit{when} and \textit{how} users need \textit{what} types of information. 
However, there is a limited understanding of the \textit{action needs} users have in-situ. 
To bridge this gap, we ran a \textcolor{black}{formative workshop} followed by a diary study to collect and identify the actions people might take when interacting with multimodal information.
In contrast to collecting and reflecting on already captured data in smartphones, \textcolor{black}{the diary study} prompted participants to \textit{actively} capture fresh data immediately, \ie the actions they intended to take whenever they encountered new multimodal information.
This approach mirrored the way users interact with information in AR settings, simulating an ``always-on'' audio-visual sensor. 
The collected data (\ie visual inputs such as scenes, physical objects, texts, and auditory inputs such as acoustic sounds, human speech) were then documented as images or text descriptions for further analysis.
\textcolor{black}{
Over the course of five days, 39 participants contributed 382 data entries.
The collected data was then used to inform the creation of a design space of possible follow-up actions that should serve as a blueprint for the design of possible follow-up actions that future interactive systems could incorporate~(Figure~\ref{fig:flowchart}e).}

\textcolor{black}{
The design space was then used to inform the design of a prototype  called \codename, containing a pipeline which enables the simultaneous processing of multimodal sensory inputs and subsequent generation of follow-up action predictions on target information (Figure \ref{fig:flowchart}f).
}
Powered by a large language model (LLM), \codename ~\textit{(1)} converts multimodal sensory inputs into structured text via existing models (\eg visual language models for image captioning) and then \textit{(2)} leverages the explicit reasoning of the LLM \cite{huang2022towards} on the structured text to \textit{(3)} predict target information (\eg the visible text) and follow-up actions (\eg share with another person) grounded in the design space (Figure \ref{fig:flowchart}g). 
\textcolor{black}{
To demonstrate the effectiveness of our pipeline and explore the LLMs' capabilities to support such real-world tasks, we conducted an evaluation using the empirical data collected from the diary study and compared multiple techniques of using LLMs. 
\begin{colortext}
We employed three variants of using LLMs: conventional intent classification, in-context learning with Chain-of-Thoughts (CoT) prompting, and fine-tuning with CoT prompting.
\end{colortext}
The results show that our approach yields competitive performance. For instance, in-context learning with CoT prompting using the latest LLM (\ie \texttt{GPT-4}) achieved a high accuracy (94.3\%) \begin{colortext}
when predicting the top three possible general actions.\end{colortext} 
As an instantiation of the pipeline, we also developed an interactive \textcolor{black}{smartphone} prototype for user interaction. We conducted an in-lab feedback session with 5 participants to collect initial subjective feedback about the system and insights to improve the design and user experiences with the interactive prototype. 
}

\textcolor{black}{
The contributions of this research are thus:}

\begin{itemize}
    
    \item \textcolor{black}{ A design space of follow-up actions that can be performed in response to multimodal sensory inputs. This design space was derived from the diary study data and surfaced 7 general and 17 specific categories of follow-up actions.
    }
    \item A novel pipeline, \codename, that provides generalized predictions of follow-up actions for real-world multimodal sensory inputs. \codename leverages the explicit reasoning of LLMs (CoT) on structured text converted from multimodal data to ground the predicted actions in the design space. 
    

    \item An evaluation of the approach enabled by empirical data collected from the diary study using different techniques (i.e., in-context learning and fine-tuning). 
    The results showed competitive performance of the proposed approach. Additionally, the evaluation provided insights into LLMs' capabilities to support real-world tasks.
    

    \item \textcolor{black}{An interactive \textcolor{black}{smartphone} prototype that predicted users' target information and suggested follow-up actions. User feedback highlighted the system's potential and the design space's comprehensiveness. }
\end{itemize}

\end{colortext}


%% file: figures/flowchart.tex
\begin{figure*}[ht!]

    \centering
    \includegraphics[width=\linewidth]{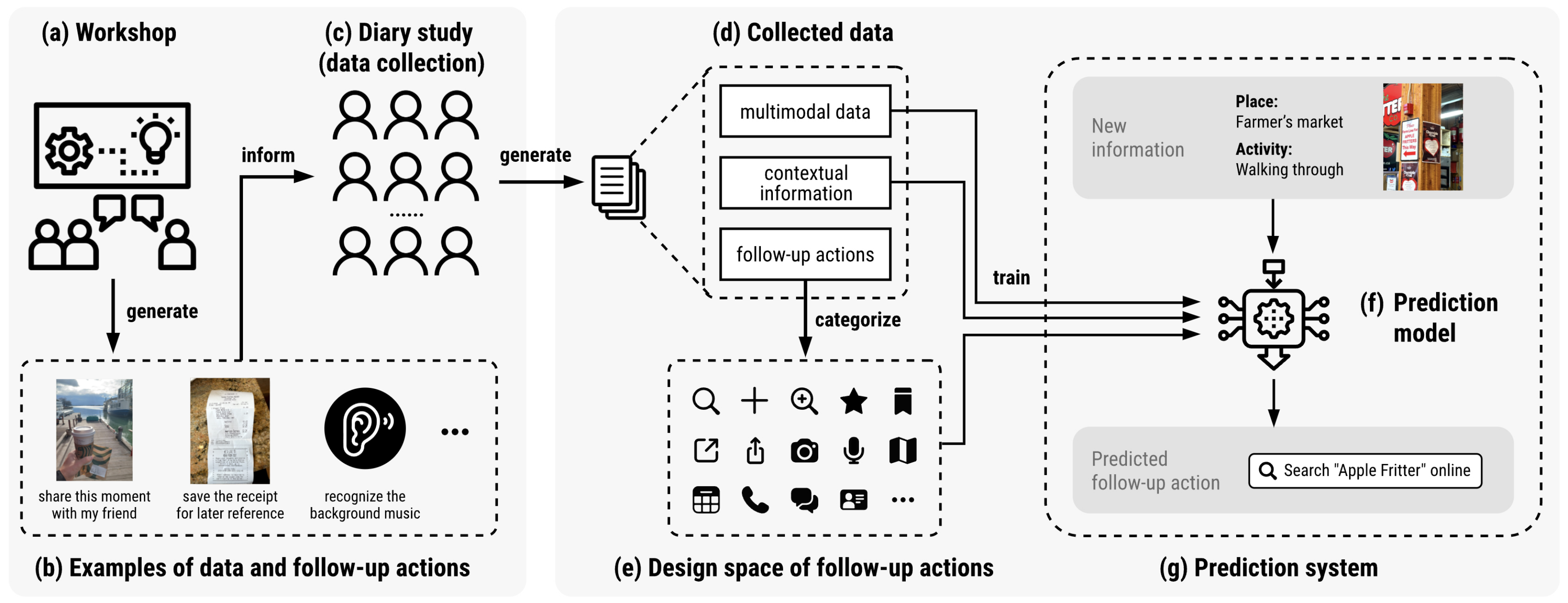}
    \caption{The development process for \codename. (a) An internal workshop was conducted to (b) generate informative examples of situations when users may take using multimodal information. (c)
The examples were used to inform and inspire the participants during a diary study that (d) collected data when participants wished to take action using multimodal data.
(e) The follow-up actions submitted by participants were then analyzed and categorized into a design space. (f) The collected data included contextual information that was used to train a prediction model that was (g) integrated within \codename to predict multiple follow-up actions given multimodal information.
}
    \label{fig:flowchart}

\end{figure*}

%% file: content/02_related.tex
\section{Related Work}
The present research was inspired by prior work on 
\begin{colortext} 
   users' mobile information needs,
\end{colortext} multimodal information interaction techniques, and the use of large language models to augment interaction. 

\begin{colortext}
\subsection{Mobile Information Needs}
Information needs, defined as "\textit{any information that is required for a task, or to satisfy the curiosity of the mind, regardless of whether the need is satisfied or not}" \cite{dearman2008examination}, is closely related to how users interact with real-world information. 
Researchers have conducted various diary studies~\cite{church2014large,dearman2008examination,sohn2008diary,church2009understanding,amin2009fancy,hinze2010contextual,chen2010diary,cherubini2011barriers,komaki2012does} to understand users' information needs under different contexts, including while using mobile phones \cite{church2009understanding, hinze2010contextual, cherubini2011barriers}, seeking information within a social network \cite{dearman2008examination}
or being on-the-go \cite{sohn2008diary,chen2010diary}.
While this presents similar use cases as what we expect to encounter in pervasive AR systems,
existing research majorly focuses on \textit{what} types of \textit{information} users require, and \textit{how} their \textit{contexts} affect their needs.
However, there is a notable gap in understanding the next phase of addressing \textit{actions needs}: \textit{what} types of \textit{actions} users might take once their information needs have been met. 
Perhaps most related is prior work by Church \etal, which explored the types of searches (\ie informational, geographical, or personal information) associated with different contexts~\cite{church2009understanding}.
The scope of these follow-up actions, however, was limited to searching for target information, rather than to a broader range of actions that could be performed with the information.
To bridge this gap, \codename aims to understand what \textit{actions} users might take once they have access to the information they need. 
We envision the potential for \codename to enable rich contextual understanding in future AR scenarios, therefore, we focus specifically on the real-world information that can be perceived by the sensors on an AR device when using different modalities.

\end{colortext}

\subsection{Multimodal Information-Based Interaction Techniques}
To predict follow-up actions while encountering new information in the wild (e.g., music, noise, visible text, objects, etc.), it is crucial that systems are able to understand the context of one's environment and the information that is available to users. One way to obtain such an understanding is to directly retrieve information that is embedded in barcodes, fiducial markers \cite{garrido2014automatic}, human faces \cite{ajanki2011augmented}, or objects during fabrication processes \cite{dogan2022infraredtags, li2017aircode, dogan2020g}. Researchers have also explored retrieving "raw" information such as visible text \cite{saluja2019ocr, zhan2019esir}, physical objects \cite{ren2015faster, girshick2015fast}, multimodal scenes \cite{zeng2022socratic}, human speech (e.g., Google API\footnote{https://cloud.google.com/speech-to-text}), and music (e.g., Shazam). Nevertheless, to understand users' intent based on the information in their physical environments, multimodal information must be monitored and processed in a way that a system can make predictions using it.

Lifelogging digitally tracks a person's daily experiences and is one way to process multimodal information \cite{gurrin2014lifelogging, ksibi2021overview}. Prior work has used lifelogging to enhance human memory by retrieving moments through natural language \cite{dubourg2016sensecam, sellen2007life} or monitor one's health by analyzing logged data \cite{lee2019development}. However, lifelogging does not specifically focus on predicting a user's intent and to predict follow-up actions, which requires the categorization of the design space. 
Several lifelogging datasets have been collected, including the Aria dataset \cite{aria_pilot_dataset}, Ego4D \cite{grauman2022ego4d}, and other video datasets \cite{rossetto2015osvc, rossetto2019v3c}. These datasets could be used to investigate desired follow-up actions, but they contain redundant data when such actions are not required. To specifically explore follow-up actions with multimodal information, we conducted a diary study prompting participants to log data whenever they wanted to act on their captured information. Building on prior research on processing multimodal information, we used this data to develop a system capable of predicting follow-up actions.

\subsection{Large Language Models in HCI}

\begin{colortext}Artificial Intelligence (AI)\end{colortext} has been widely used in the Human-Computer Interaction (HCI) community, with LLMs experiencing a surge of usage in recent years \cite{ahn2022can, kirilyuk2023visual, park2022social, petridis2023anglekindling, dang2023choice, jakesch2023co, wang2022enabling, hamalainen2023evaluating, wang2021popblends, jo2023understanding, Wang2024LAVELA}. LLMs' abilities to understand common knowledge and reason within a given context have been leveraged for interactive code support \cite{wang2022enabling}, social computing \cite{park2022social, park2023generative} and accessibility support \cite{huh2023genassist}. For example, Visual Caption employed a fine-tuned language model to predict user intent during visual inquiries using the last two sentences \cite{liu2023visual}, while SayCan extracted and leveraged knowledge priors within LLMs to reason about, and execute, robot commands \cite{ahn2022can}.
LLMs have also been used to enhance recommender systems that utilize contextual information to recommend items \cite{burke2007hybrid, klouche2015designing}. For example, GPT-3 \cite{brown2020language} was used to augment movie recommendation systems \cite{zhang2021language}.

Most of this prior work relied on the capture of one's explicit intent \cite{chen2023next}, wherein users or agents interacted with a system via direct prompts. \codename{} unlocks a new interaction method with LLMs by embracing a more implicit intent, focused on the user's current visual input (e.g., multimodal information such as environmental understanding or recognized text) and contextual information. 
Coupled with the Chain-of-Thoughts prompting, this enables \codename{} to deliver explainable predictions of target information and follow-up actions.

%% file: content/03_formative_study.tex
\section{Formative Workshop}
\label{sec:formative}

%

We ran a formative workshop to obtain a preliminary understanding about the multimodal information triggers people came across in everyday life and their follow-up actions. The outcomes of the workshop were clusters of the actions participants took with multimodal information triggers. The learnings on the workshop informed our method choices, question design, and example generation for the next study to collect data from general population.   




\subsection{Procedure}
We recruited 10 participants within our institution through group email invitations.
The participants included HCI researchers, UX designers, and student interns, all of whom worked within the domain of AR and XR.
\textcolor{black}{Their expertise would provide insights on how people may interact with information in the physical world.}
The participants volunteered to join the workshop and they were not paid. The workshop consisted of three parts and lasted one hour in total. Participants were invited to use a FigJam\footnote{https://www.figma.com/figjam/} whiteboard for synchronous collaboration.

\begin{figure}[]
    \centering
    \includegraphics[width=\linewidth]{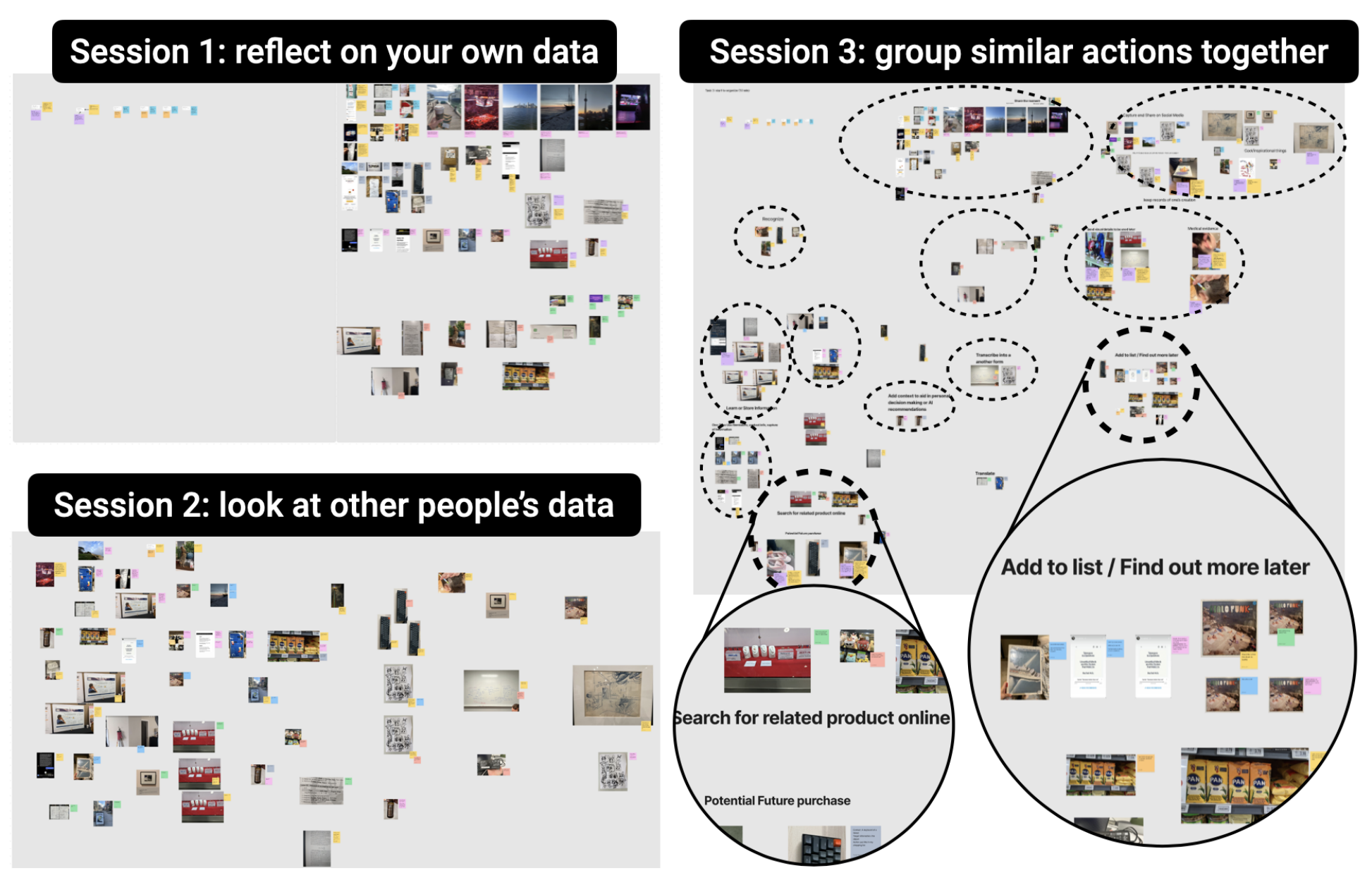}
    \caption{Screenshots from the formative workshop where participants shared data in Session 1, reviewed other participants' data in Session 2, and grouped similar actions in Session 3.}
    \label{fig:workshop_figma}
\end{figure}

\subsection{Process}
The organizer of the workshop first introduced the goal and agenda of the workshop to the participants. Then they shared two examples with the participants, including a parking ticket and an audio file of some background music, and their related context and follow up actions. During Part 1, participants were asked to share their own media, context, and follow-up actions. During Part 2, participants reflected on other participants' media and came up with their follow-up actions. In Part 3, participants collaboratively clustered similar actions (Figure~\ref{fig:workshop_figma}).
 

\subsubsection{Part One}
\textit{``Browse past media, share those that you did follow-up actions with''.}
During this part, participants had 20 minutes to browse their personal media storage and upload the ones that they took actions with to shared Google drive and the FigJam board.
For each shared media item, participants were asked to recall the record the following information:
\begin{colortext}
\one what target they acted on (\eg the menu of a boba shop),
\two what action they took (\eg save to the album for future reference) and 
\three contextual information such as the location or their activity, which is useful in the next part.\end{colortext}
For audio and video uploads, the participants described them textually on the FigJam board. 
Participants shared a total of 66 examples (i.e., 6 video/audio clips and 60 images) and 66 follow-up actions.

\subsubsection{Part Two}
\textit{`` Imagine if you were the person at the scene, what actions you would take on the information?''}
In this part, we aimed to get third-person perspective on what the possible actions could be given the media. 
\begin{colortext}Contextual information from part one helps other participants to imagine the scenarios.\end{colortext} Participants had 20 minutes to browse examples shared by other participants and to type their imagined follow-up actions for the target information. 
An additional 104 follow-up actions were proposed, with a total of 170 follow-up actions between session one and two.

\subsubsection{Part Three}
\textit{``Now group together those actions that are similar.''}
In Part 3, participants had 15 minutes to collaboratively cluster and label all 170 examples from Part 2, using an affinity diagram.

\subsection{Results}
\label{sec:initial_designspace}
After the workshop, two researchers coded, filtered, and clustered the 170 follow-up actions independently. 
\begin{colortext}
The participant-generated clusters were also referenced in this process. This process was inductive, meaning that they coded actions mentioned by the participants, rather than starting with an existing set of actions.
The results from each researcher and the clusters from participants were compared. The researchers discussed and resolved the discrepancies in the clusters' boundary, naming, and granularity.\end{colortext} As a result, they identified 13 types of actions that were grouped into four categories (i.e., share, save, query, and others; Figure \ref{fig:initial_space}). Representative examples from these categories were used as learning materials for participants in our subsequent diary study.

\input{figures/initial_designspace}


\begin{colortext}
One important observation was that participants seldom captured or shared audio. This might be due to the fact that audio contains temporal information that is hard to capture (\eg an abnormal sound that occurs intermittently).\end{colortext}
This finding informed the design of the diary study, where we asked participants to share the textual description of their audio rather than the audio itself. We present more details in the next section.

%% file: figures/initial_designspace.tex

\begin{figure}[]
    \centering
    \includegraphics[width=\linewidth]{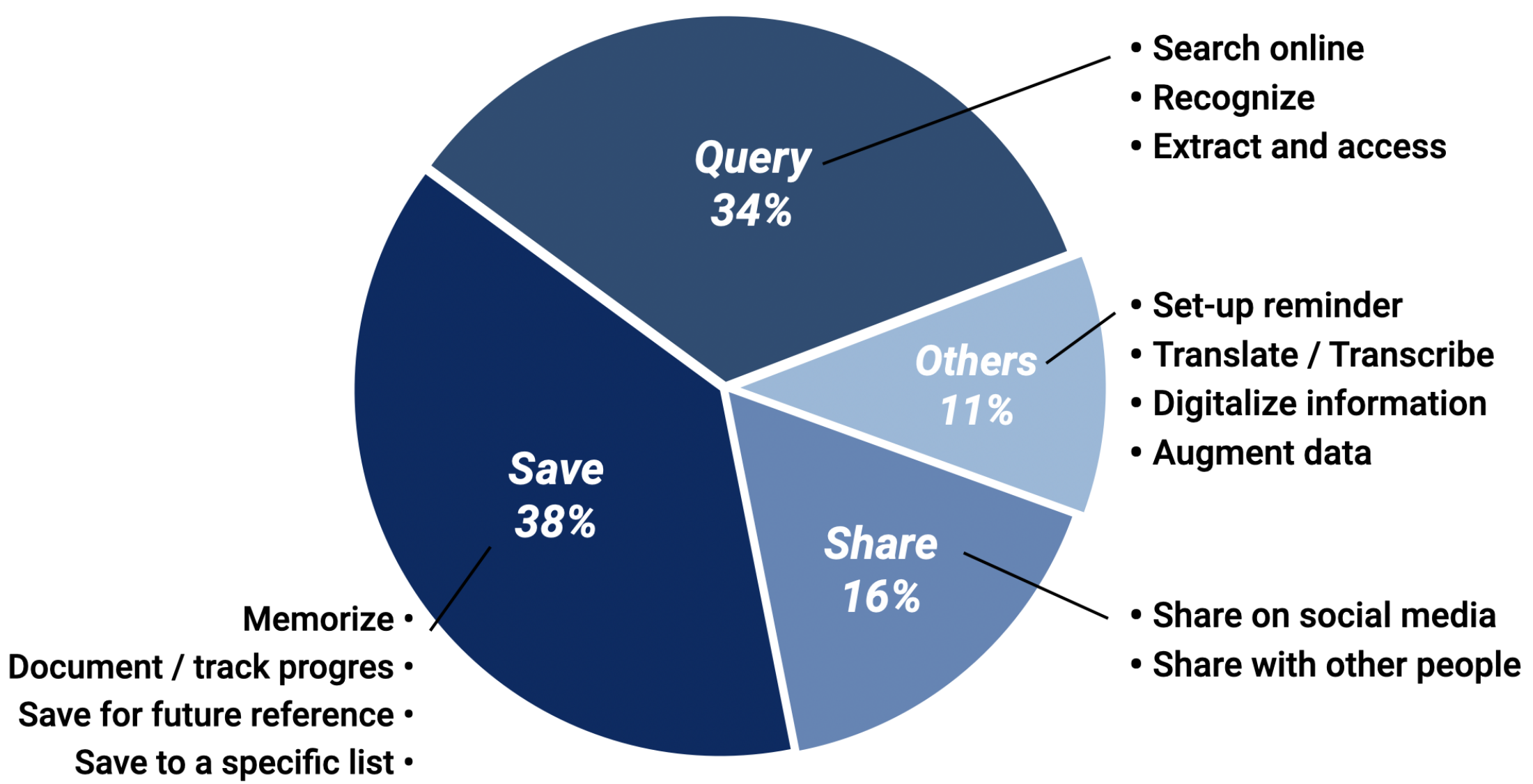}
    \caption{Frequencies of the 13 follow-up actions generated during the workshop (n = 170) that were grouped into 4 categories.}
    \label{fig:initial_space}
\end{figure}

%% file: content/04_diary_study.tex
\section{Data Collection via a Diary Study}
\label{sec:diary}

While the workshop provided an initial glimpse of the type of multimodal information and follow-up actions users would desire, we wanted to formalize the findings with in-situ experiences from participants external to our institution. The use of a diary study methodology would enable participants to log data whenever needs arose \cite{sohn2008diary}, making it an ideal choice to examine desired follow-up actions when one encounters new information. We leveraged this methodology to answer the following research question:
\begin{itemize}
    \item[\textbf{RQ:}] What follow-up actions do general users wish to take when they encounter new multimodal information in a real-world environment? 
\end{itemize}

\begin{colortext}
We adopted the \textit{snipped-based diary technique} proposed by Brandt \etal \cite{brandt2007txt} to collect data about users' follow-up actions with multimodal information.
As opposed to reflecting on captured data (\eg images in the album) at a fixed time of day, our participants were asked to log data whenever they encountered information in the world they wished to take action upon.
This simulates the ``always-on'' feature of an AR platform where users can interact with AR interfaces anytime and anywhere.
We collected the data including \one the target information they wished to take action on, \two the desired follow-up actions and \three contextual information such as their goals, locations, and activities.\end{colortext}
Contextual information was important to collect as it could affect the choice of follow-up actions \cite{schilit1994context, chen2000survey, lindlbauer2019context}. For example, looking at a shampoo bottle in a drug store has a different desired follow-up action than looking at the same bottle at home (\eg comparing the price to a similar product versus ordering another bottle on Amazon). Therefore, we hypothesized that contextual information would increase a system's ability to accurately understand users' goals and follow-up actions. We incorporated this information into a predictive model later on in our research process.

\subsection{Participants}
Thirty-nine participants (i.e., 16 male, 22 female, and 1 non-binary) were recruited from the dscout user research platform\footnote{https://dscout.com/}. All  participants were between the ages of 18 to 69 years old, were proficient in English, and had a smartphone to take photos. 
Each participant was compensated \$50 USD after they completed the diary study for their time. 


\subsection{Procedure}
\label{sec:procedure}

The diary study consisted of two phases, i.e., an introductory phase and a diary phase. 
During the introductory phase, participants were shown examples from the workshop that represented several of the categories of media and actions that the workshop participants identified.
Note that to avoid bias due to the categorization that resulted from the workshop, participants were only shown the exemplar media and follow-up actions. 

\begin{figure}[t]
    \centering
    \includegraphics[width=\linewidth]{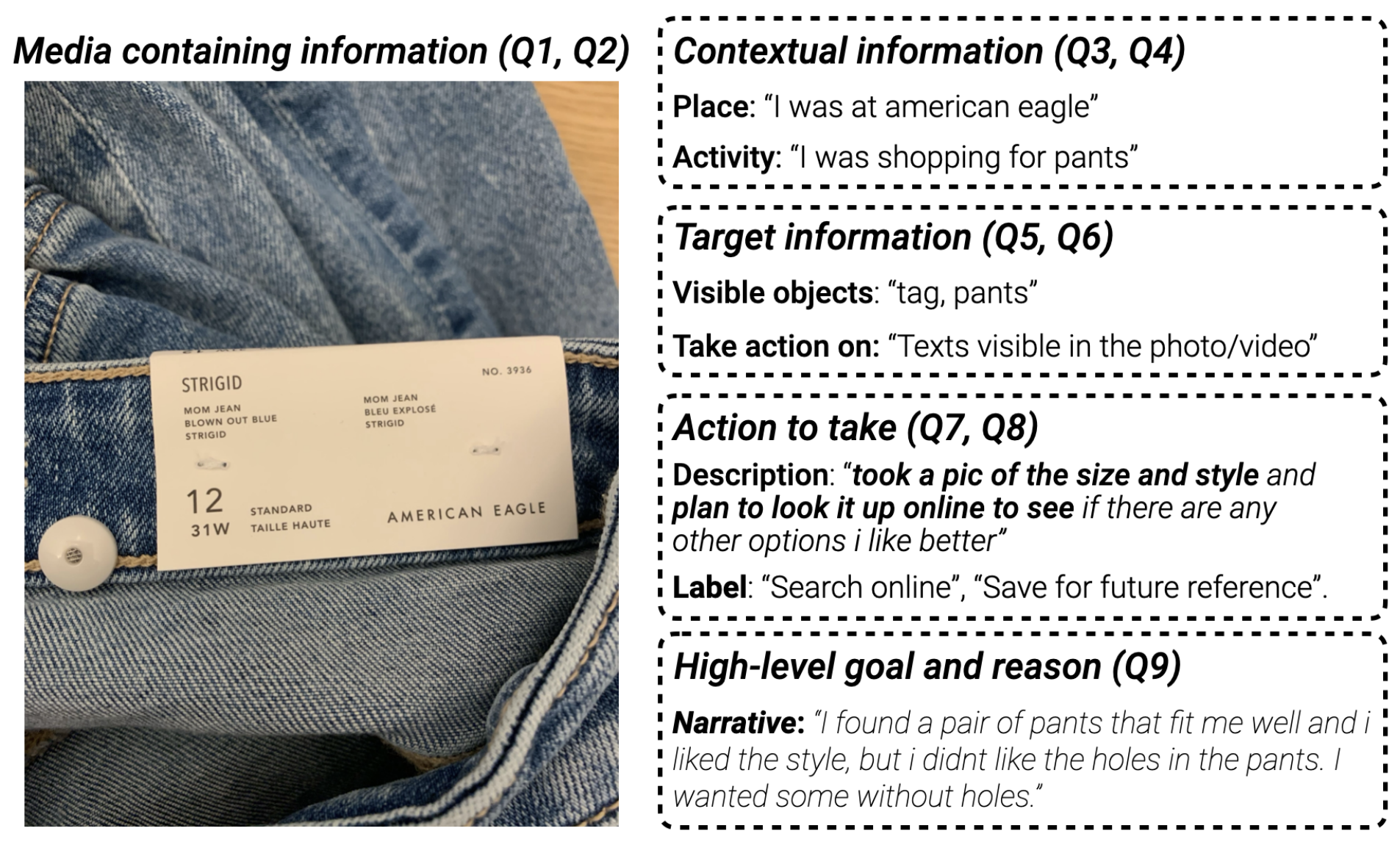}
    \caption{An example diary entry from the diary study.}
    \label{fig:example_data}
\end{figure}

During the diary phase, participants were instructed to submit 2 entries each day for five days. These entries needed to reflect genuine participant needs that occurred in the moment. The diary phase began in the middle of the week and extended over the weekend to capture the different types of needs that may occur throughout a week. For each diary entry, participants were required to answer questions about their entry (Figure \ref{fig:example_data}). These included: 

\paragraph{Media Containing the Information (Q1, Q2)}
Although we aimed to collect multimodal information, we were not allowed to collect audio or video data from participants that could contain potentially identifiable personal information due to the legal requirements of our institution. Therefore, if participants wanted to share audio or video, 
they were asked to provide a text description of the data or screenshots of the videos instead (\eg ``This is the background music I heard in the cafe''). 

\paragraph{Contextual Information (Q3, Q4)}
Context was first introduced by Schilit \etal as \textit{``locations, identities of nearby people and objects, and changes to those objects''} \cite{schilit1994disseminating}.
To predict follow-up actions, we identified how the location and the user's activity would affect how users would interact with the encountered information. 

\paragraph{Target Information (Q5, Q6)}
Since we were investigating follow-up actions for multimodal information, it was essential to know which information the participant wanted to perform follow-up actions for. For example, participants could be interested in only the text visible in an image or the entire scene.
Participants were thus asked to identify
the objects visible in the image or the sounds that could be heard (Q5). This provided additional context to achieve a better understanding of potential user interactions with the information.

\paragraph{Actions to be Taken (Q7, Q8)}
Participants were asked to use natural language to describe the actions they intended to take and then categorize these actions. Additionally, they could select  categories corresponding to these actions using the action categories identified in the workshop. Participants also had the option to create new categories by selecting 'other' if there were actions that did not fit within the existing categories. 
Note that we minimized potential bias by asking participants to detail their intention and desired actions in their own words on a first page before being shown and asked to choose from the action types on the next page. Participants selected categories that were later used as a reference point during the iteration towards the final design space presented in the following sections. 

\paragraph{High-Level Goal and Reasons (Q9)}
To better understand why participants intended to take certain follow-up actions, we asked participants to share their high-level goals and reasons for doing so. 



\subsection{Data Summary}

During the study, \begin{colortext}two participants did not finish the number of required data entries (one only submitted 7 and the other only 5) and they were compensated \$5 per submitted entry\end{colortext}. This resulted in 382 data entries in total.
The ratio of collected visual to audio data was approximately 2:1. 
We collected 254 visual data examples (i.e., 193 photos and 61 videos with visual selected as the target information type) and 128 audio data examples (i.e., 48 videos with audio as the target information type and 80 text descriptions of audio). Participants reported wanting to take action on 55 full scenes \begin{colortext}(40 photos / 15 videos)\end{colortext}, 120 individual objects \begin{colortext}(96 photos / 24 vidoes)\end{colortext}, 79 pieces of text \begin{colortext}(57 photos / 22 videos)\end{colortext}, 51 speech clips \begin{colortext}(20 videos / 31 audio only)\end{colortext}, and 77 sound clips \begin{colortext}(28 videos / 49 audio only)\end{colortext}. 
Additionally, participants shared 17 (i.e., 10 visual, 7 audio) follow-up actions which did not fit within any of the categories identified during the workshop.


\begin{figure*}
    \centering
    \begin{subfigure}[t]{0.5\linewidth}
        \centering
        \includegraphics[width=\textwidth]{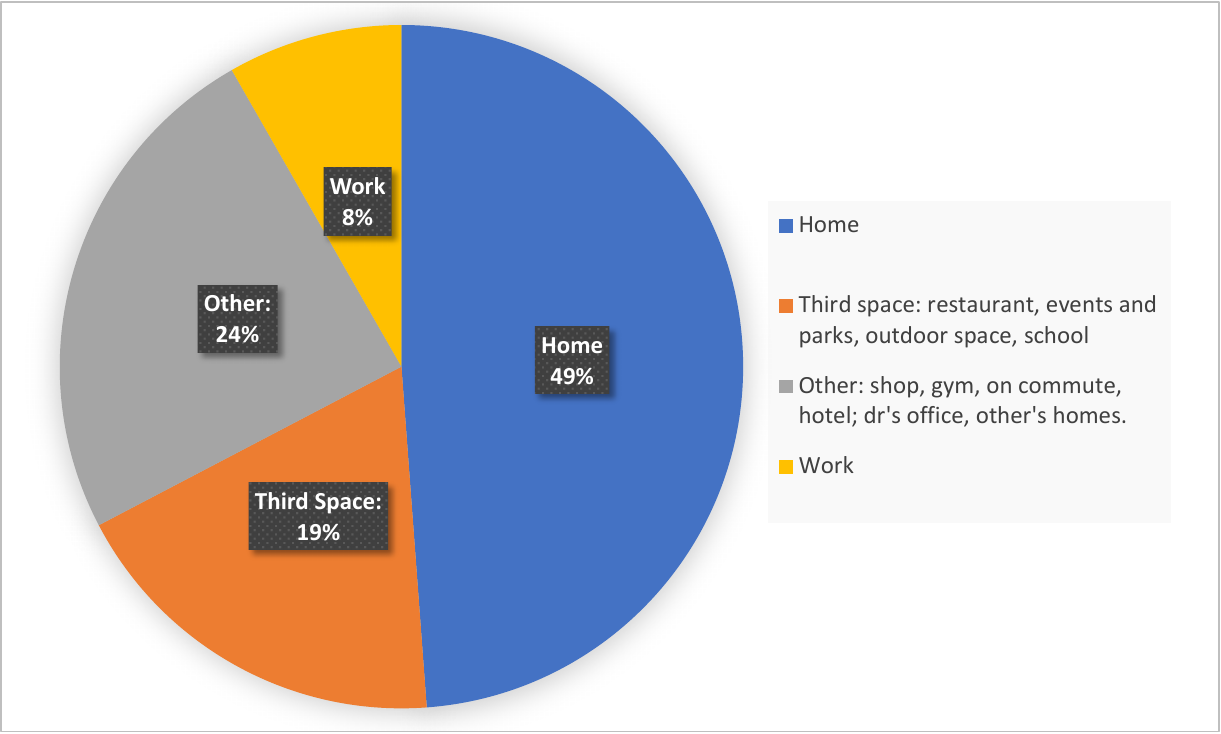}
        \caption{The location distribution of our dataset.}
        \label{fig:context_location}
    \end{subfigure}
    ~
    \begin{subfigure}[t]{0.5\linewidth}
        \centering
        \includegraphics[width=\textwidth]{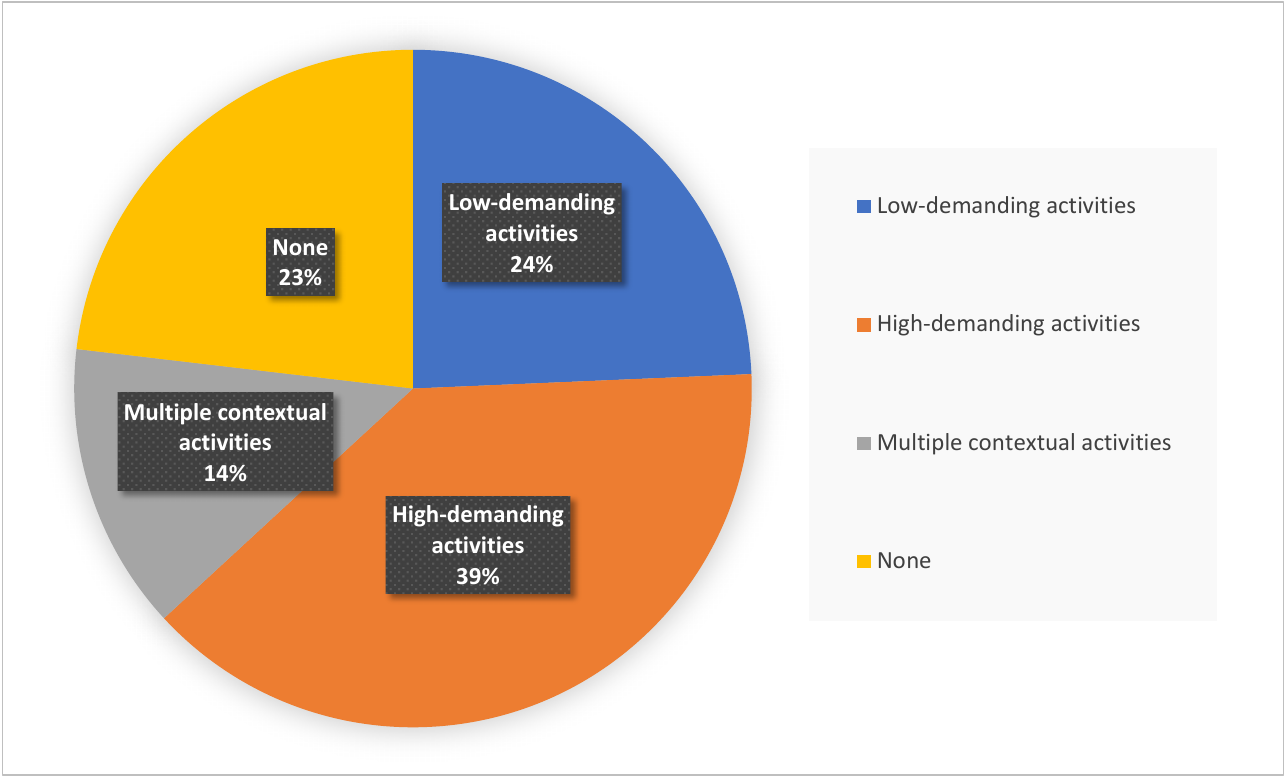}
        \caption{The activity distribution of our dataset.}
        \label{fig:context_activity}
    \end{subfigure}
    
    \caption{\begin{colortext}
In (a), third space refers to the places outside of home or work where people have the potential opportunity to socialize and engage with the community \cite{oldenburg1999great}.
    In (b), the low-demanding activities include: Sedentary leisure activities (i.e. watching TV, browsing social media, browsing news, drawing, reading), Eating/drinking, Waiting, Sedentary housework (i.e. checking emails, online payments, online shopping, personal care); The high-demanding activities include: Interacting with someone, Physical housework (i.e. cleaning, cooking, organizing, maintaining, getting mails, gardening), Full-body movement activities (i.e. walking, working out, playing), Focused activities (i.e. driving, studying, working), Shopping in a store, Preparing with time pressure, Exploring and navigating environment.        
    \end{colortext}}
    \label{fig:context}
\end{figure*}



\begin{colortext}
\subsubsection{Contexts of the Captured Data}
We coded and summarized the contexts when people came across multimodal information that they intended to take follow-up actions based on survey answers in Q3 and Q4. Figure \ref{fig:context} shows the diversity of location and contextual activities people had. 
We consider our dataset to be representative to a day in the life, based on the comparison to the American Time Use Survey (ATUS, from U.S. Bureau of Labor Statistics) \cite{ATUS}. The diversity of the contextual activities included all activity categories mentioned in the 2022 ATUS survey \cite{ATUS} except "sleeping" (not applicable to our study), "caring for non-household members", or "organizational, civic, and religious activities". The latter two categories together accounted for 0.5 hours per day per person on average. 
Most (77\%) of the in-situ capture about people's follow-up actions needed had other contextual activities, out of which 24\% were low-demanding activities and 39\% were high-demanding activities that require full body motion or high cognitive focus, and 13\% involved both types of contextual activities. This showed the pervasiveness of multitasking situations where people's physical and cognitive bandwidth were already used for other activities. Therefore, it was important to reduce the friction for people to use follow-up actions.

\end{colortext}


%% file: content/05_design_space.tex

\input{figures/designspace_figure}

\section{Design Space of Follow-up Actions}
\begin{colortext}
Following the diary study, a researcher and research assistant collaboratively reviewed the diary entries, coded the data, and compared and consolidated the codes through iterations.\end{colortext}
The resulting action space consisted of 7 \textbf{general categories} of follow-up actions, including \textit{share}, \textit{save}, \textit{remind}, \textit{look up}, \textit{digital extract}, \textit{media manipulation}, and \textit{complex actions}. These categories were further divided into 17 \textbf{specific categories} (Figure \ref{fig:designspace}). 

\begin{colortext}
For the general categories, (1) \textit{Share} refers to actions that users employ to make information available to others (i.e., sending information to friends or family or posting the information on a social media platform such as Instagram or Facebook); 
(2) \textit{Save} refers to the actions used to store information;
(3) \textit{Remind} refers to actions that created an alert or notice to remember something later such as setting a reminder after seeing a flight schedule on a screen or noting oneself of the date of a specific event (particularly useful for managing tasks, appointments, or important events);
(4) \textit{Look up} refers to actions that searched for specific information or details;
(5) \textit{Digital extract} refers to actions taken to obtain and utilize information from multiple sources;
(6) \textit{Media manipulation} refers to actions that altered or modified media content to achieve a specific outcome, and
(7) \textit{Complex actions} involve processing data from multiple sources.
Figure~\ref{fig:designspace} lists the definition of the 17 specific categories; please refer to Appendix \ref{sec:category} for more detailed explanation.
\end{colortext}

\begin{figure}[t]
    \centering
    \includegraphics[width=1\columnwidth]{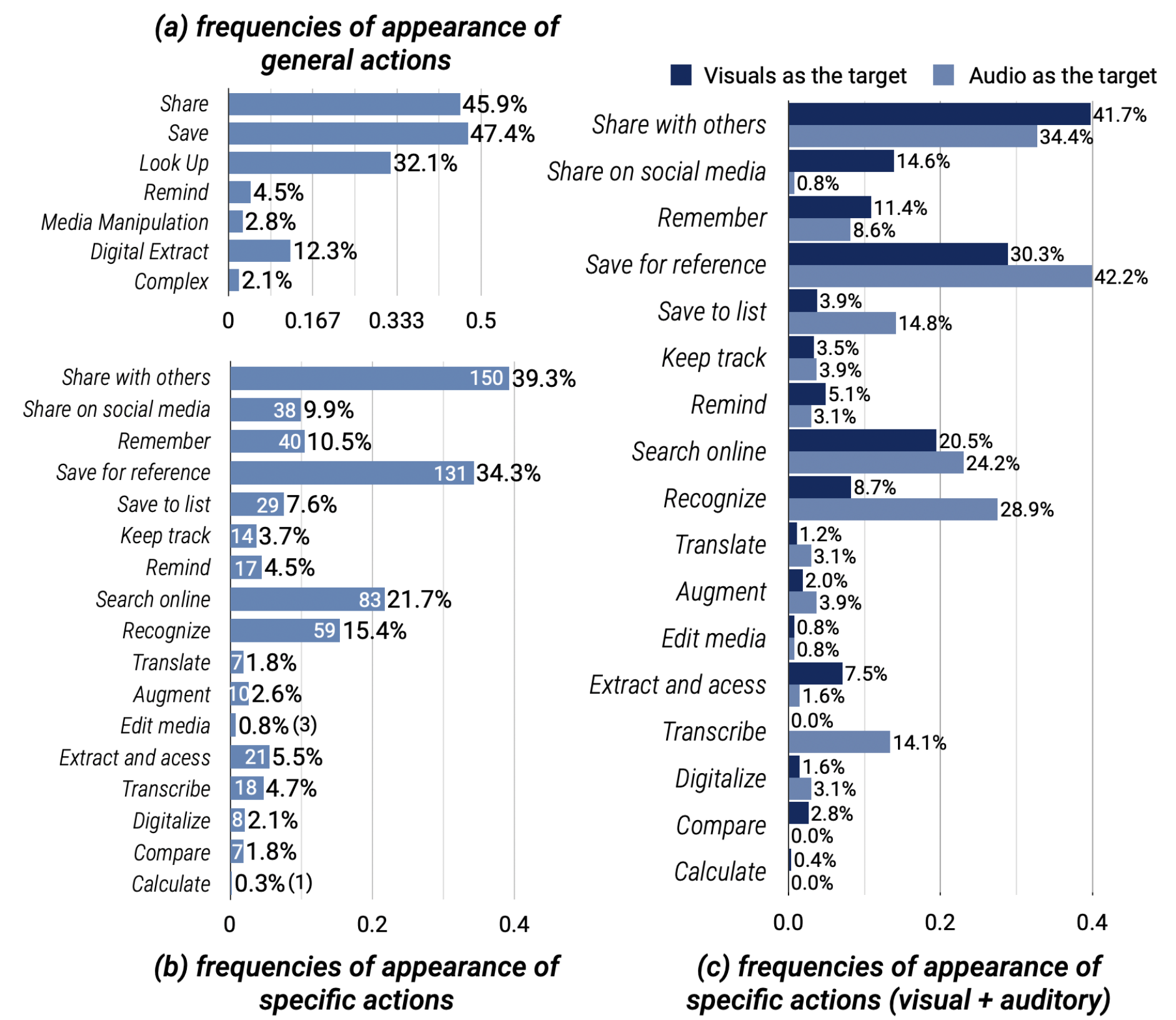}
    \caption{\begin{colortext} (a) The frequencies of the general actions. \end{colortext}(b) The frequencies of the specific actions (with number). (c) The frequencies of the specific actions on visual and audio. Frequency was computed as the number of appearances divided by the total number of diary entries.}
    \label{fig:action_dist}
\end{figure}

\subsection{Analysis of Diary Data Using the Design Space}
\label{sec:data_analysis}

We conducted a post-study analysis on the diary study data using the categories within the design space (Figure \ref{fig:action_dist}). \begin{colortext}
The \textit{share} (45.9\%), \textit{save} (47.4\%), and \textit{look up} (32.1\%) actions were most common general actions 
while the remainder of the actions (i.e., \textit{remind} (4.5\%), \textit{media manipulate} (2.8\%), \textit{digital extract} (12.3\%), \textit{complex actions} (2.1\%)) were less common (Figure \ref{fig:action_dist}a). \end{colortext} 
Figure \ref{fig:action_dist}b shows the frequencies of each specific action.
Within the data, we also observed that participants tend to take multiple actions in succession. For example, participants \textit{remembered} a memorable moment and then \textit{shared} it with family members.
\begin{colortext}
Specifically, 183 diary entries had only one action, 147 had two actions, 44 had three actions and 8 had four actions.
An example with four aggregated specific actions is illustrated in Appendix \ref{app:aggregated}.
\end{colortext}


Participants also used different patterns of follow-up actions when interacting with data from different modalities (Figure \ref{fig:action_dist}c). The overall frequency of specific follow-up actions when the target was visual versus audio were similar, although there appears to be a difference when \textit{sharing on social media}, \textit{saving to a list}, \textit{recognizing} and \textit{transcribing}.
These variations align with typical real-world interactions.
\begin{colortext}
For example, people often share visual content (\eg a breathtaking landscape or an unusual statue) on social media, while it is less common to post specific sounds (\eg an abnormal noise) in an environment.
\end{colortext}
Additionally, as described earlier, \textit{transcribing} is exclusive to audio.
Furthermore, our data showed a trend where individuals \textit{recognized} and \textit{saved} music to their playlists upon hearing a song they enjoyed. This reflected how people interact with, and respond to, real-world audio data, 
\begin{colortext}
which also leads to a higher frequency of \textit{saving for reference} actions in similar scenarios.
\end{colortext}

\input{content/06_prediction_model}



%% file: figures/designspace_figure.tex

\begin{figure*}
    \centering
    \includegraphics[width=\linewidth]{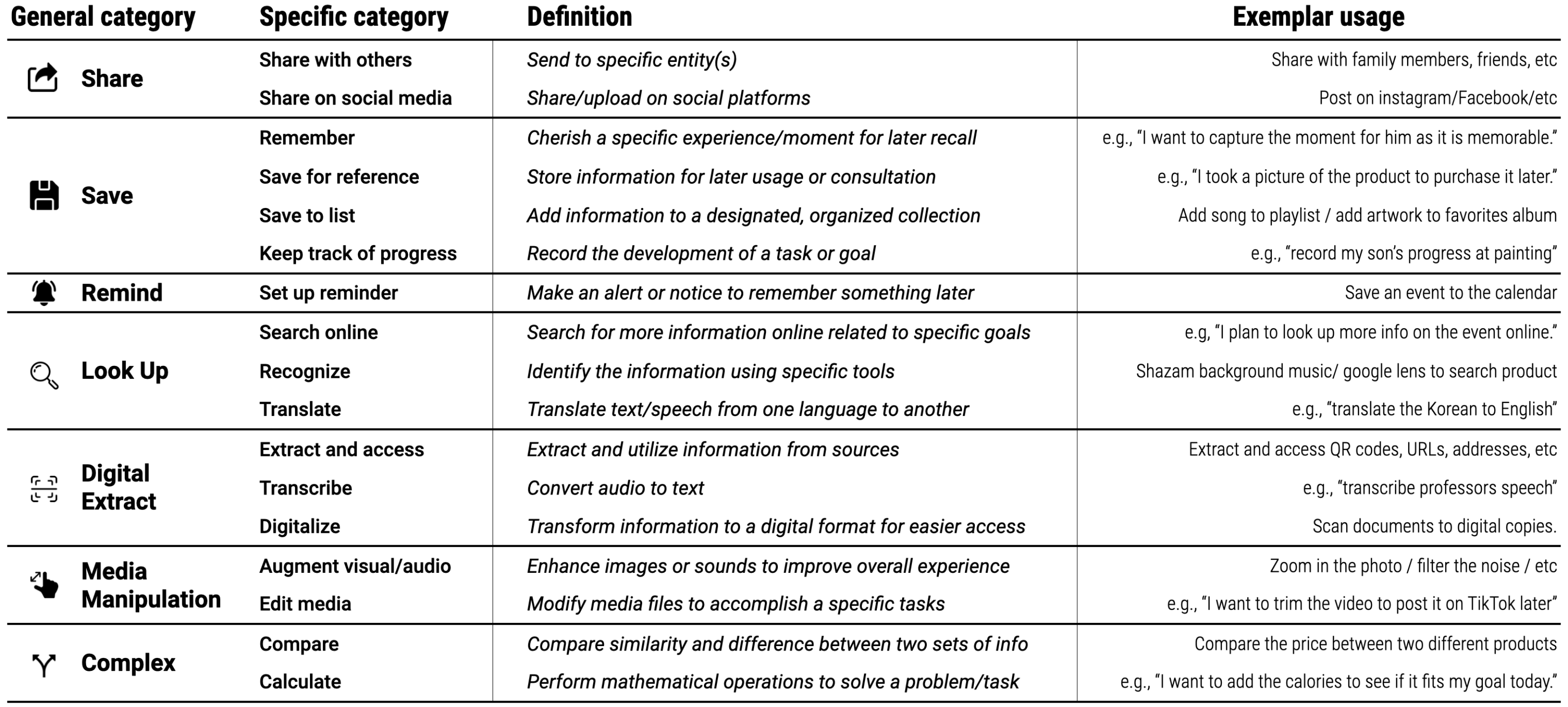}
    \caption{Design space of follow-up actions for multimodal information that emphasizes general and specific categories of actions. 
    }
    \label{fig:designspace}
  \end{figure*}

%% file: content/06_prediction_model.tex
\section{\codename Pipeline}

\label{sec:model}

\begin{colortext}

To reduce users' frictions to access follow-up actions triggered by the multimodal information in the world, we create \codename. The pipeline of \codename senses and processes different multimodal information, and predicts the \textit{target information} and \textit{follow-up actions} grounded in the action space, which is based on the empirical data. 
Moreover, by reasoning with multimodal and contextual information, this pipeline aims to enhance explainability and model performance.

\begin{figure*}[t]
    \centering
    \includegraphics[width=\linewidth]{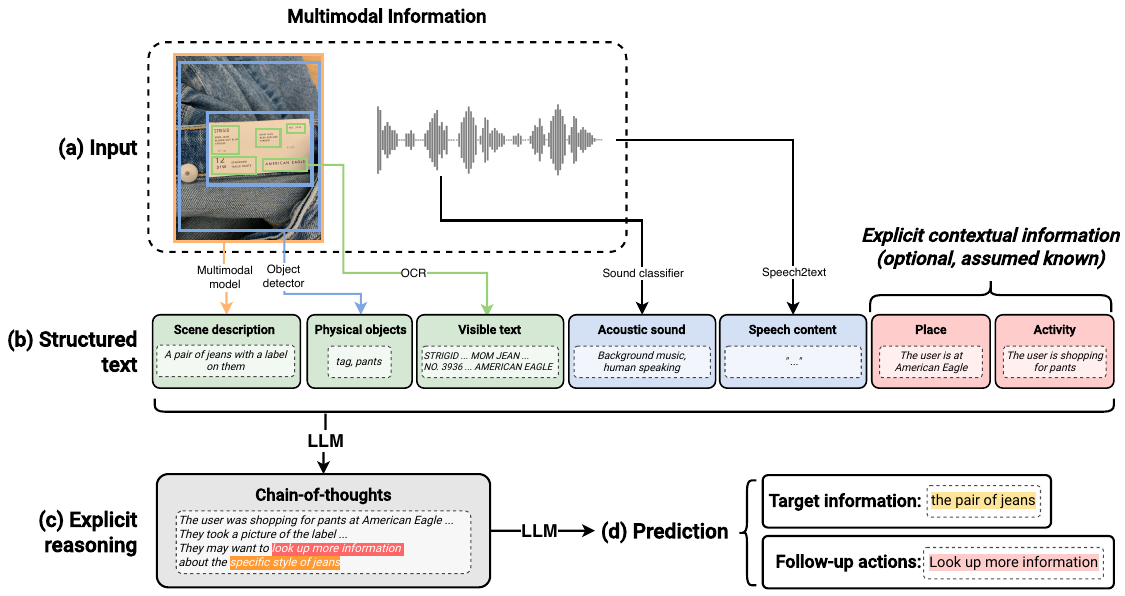}
    \caption{\begin{colortext} \codename processes multimodal information (a) by converting it into structured text using existing models (b). It processes visual data using multimodal models, object detectors, and OCR models and processes audio data via sound classifiers and speech-to-text models. Then, \codename performs an explicit reasoning using Chain-of-Thoughts prompting (c) and predicts target information and follow-up actions (d). \end{colortext}}
    \label{fig:technical}
\end{figure*}

To achieve this, \codename consists of three steps (Figure \ref{fig:technical}):
\begin{enumerate}
    \item \codename converts raw multimodal data (i.e., visual and audio data) into structured text by leveraging existing models.
    \item \codename then performs intermediate explicit reasoning on the structured text via Chain-of-Thoughts (CoT) prompting. The training data for this prompting was grounded in the data from the diary study.
    \item Finally, \codename predicts the \textit{target information} (i.e., the whole scene, physical objects, text, sounds, or speech) and the \textit{follow-up actions} grounded in the design space using a large language model (LLM). 
\end{enumerate}



\subsection{Converting Multimodal Data into Structured Text}
For a model to process information in multiple modalities simultaneously and perform predictions, it is essential to convert the multimodal data into a unified representation format (\eg a textual representative or a joint embedding space). 
This would enable a model to identify and learn from patterns in the multimodal input.
To enable explicit reasoning for prediction, \codename converted multimodal data into a textual representation. 
Specifically, \codename leveraged existing models to convert both visual and audio data into structured text before performing CoT prompting-based reasoning steps. 
All the converted data for each entry was stored in JSON format for explicit reasoning.
Note that our pipeline aims to demonstrate  potential using currently available data and could be adapted to broader range of modalities in the future.


\subsubsection{Visual Information}
Aligning with the findings from our diary study, \codename supports three aspects of visual information: the overall scene, physical objects, and any visible text.  For the overall scene, \codename leverages recent advancements in multimodal learning frameworks that have shown competitive performance in describing a scene with text. 
In this implementation, we used an open-source, state-of-the-art image captioning model, InstructBLIP \cite{dai2023instructblip}, with the prompt of \textit{``Write a short description for the image.''}. For the physical objects and visible text, \codename used the Detectron2 object detection model \cite{wu2019detectron2} to detect the objects and Google Cloud Vision\footnote{https://cloud.google.com/vision/docs/ocr}) to recognize the text.

\subsubsection{Audio Information}
\codename classified the type of acoustic sounds via YAMNet\footnote{https://github.com/tensorflow/models/tree/master/research/ audioset/yamnet} and used speech-to-text models to transcribe human speech. 
As our institution would not permit the collection of personal identifiable information, we were unable to collect human speech data during the diary study. As a result, the evaluation of our model's capabilities does not incorporate transcribed speech.

\subsubsection{Explicit Contextual Information}
As context affects the actions people perform using the information they have available to them, \codename leveraged the data collected during the diary study, i.e., where participants were and what were they doing when encountering the information.
However, such contextual information may not always be available in practice, and thus this is optional to include in our pipeline.
We examined the impact of the contextual information on the prediction performance in Sec. \ref{sec:ablation}.

\end{colortext}

\subsection{Generating Chain-of-Thoughts Prompts}
Traditional classification methods typically rely on trained models like black boxes. To enhance explainability, a model should explain the rationale behind its predictions for certain follow-up actions. Ideally, this reasoning should be as close to a user's reasoning as possible. This is especially important when there are multiple actionable information items captured and the user's intention is not clear from the sensor data itself. For example, in Figure \ref{fig:example_data}, the person captured an image with multiple texts (including the brand name, the jean's name and the size etc.), but the user only intends to search more information about the specific jean's sizes, rather than the brand name. Such reasoning could be instrumental for subsequent interactions, such as deciding which target information to search.
\codename addresses this by introducing CoT prompting \cite{wei2022chain} as an intermediate reasoning step through the prompting and training process (Figure~\ref{fig:technical}c). 

\begin{colortext}

One of the challenges is the generation of CoT prompts. Prior work mostly leveraged zero-shot prompting (i.e., using prompts such as \textit{"let's think step-by-step"}) or researcher-crafted prompts for in-context learning. 
However, these approaches rely on either common sense reasoning or researcher reasoning, which may not represent how our participants reasoned within their context. 

To address this, we leveraged the data collected during the diary study to generate CoT prompts in empricial data.
During the diary study, we collected participants' high-level goals and reasons (Sec. \ref{sec:procedure} (Q9)) to understand the rationale behind their intended follow-up actions. We convert these reasoning from first-person perspective to third-person perspective for the CoT prompts. 
\end{colortext}
In the above example, the participant shared their reasoning in the survey (Figure \ref{fig:example_data}):
\begin{quote}
    \textit{``I found a pair of pants that fit me well and I liked the style, but I didn't like the holes in the pants. I wanted some without holes. So I took a pic of \textbf{the size and style} and plan to \textbf{look it up online} to see if there are any other options I like better.''}
\end{quote}
The above data were used to generate the CoT reasoning as follows:
\begin{quote}
   \textit{ ``The user was shopping for pants at American Eagle and found a pair they might like. They took a picture of the label, which includes the style and size of the jeans. They may want to \textbf{look up more information} about \textbf{the specific style of jeans}, such as reviews or other colors available.''}
\end{quote}

We prompted the LLM to generate the CoT prompts for the model as the ground truth label for each data point collected during the diary study. Specifically, the prompt consisted of the list of actions with the respective description (Figure \ref{fig:designspace}) ground truth action label and the participants' responses for their goals and reasons. The template used to generate the CoT prompts is in Appendix \ref{app:cot}.

\section{Identify the Best performant LLM technique}
\label{sec:finetune}

\subsection{LLM Techniques and Implementation}
\begin{colortext}

With the \codename pipeline, we aim to predict the intended action on multimodal information. 
Recent LLMs' advancements has shown various techniques' competitiveness for new tasks, such as in-context learning and fine-tuning. 
To identify the best-performant among the state-of-the-art LLM techniques for \codename and draw insights in exploring LLMs' capabilities in addressing the target task,
we use the empirical data collected from the diary study to evaluate the performance of the pipeline using different techniques.


\textcolor{black}{
Specifically, we employed three different LLM techniques to predict the intended actions: 
\one intent classification, 
\two in-context learning with chain-of-thoughts prompting, and
\three fine-tuning with chain-of-thoughts training data.
We first discuss the rationale for choosing these methods and then explain them in detail.}

    
    




\subsubsection{Conventional Intent Classifier}

Prior research in Natural Language Processing (NLP) has explored numerous methods of classifying text-based data for different tasks, including intent classification~\cite{liu2016attention} or sentiment analysis~\cite{pang2002thumbs}. 
One key advantage of this is the potential use of smaller models (\eg BERT \cite{devlin2019bert} or LSTM \cite{ravuri2015recurrent}) for lower cost and faster execution.

To maintain consistency in our comparison, we fine-tuned a pre-trained LLM (\texttt{davinci} from OpenAI) to perform the intent classification. 
The \texttt{davinci} model has a smaller size compared to other GPT-3.5 models that support fine-tuning and it outputs \texttt{logprobs}, which provide confidence scores for different action predictions, enabling us to rank the top-n likely actions, similar to traditional classification models.
As shown in Figure~\ref{fig:techniques}, to prepare the training data we formatted the structured text into a tuple as the input for each data entry and use the target label (i.e., target information or the action) as the output.
We then used this data to fine-tune the LLM in the legacy prompt-completion\footnote{https://platform.openai.com/docs/guides/legacy-fine-tuning} way.
Specifically, we used 75\% of the data entries from the diary study for training and the rest for evaluation.

\begin{figure*}[t]
    \centering
    \includegraphics[width=\linewidth]{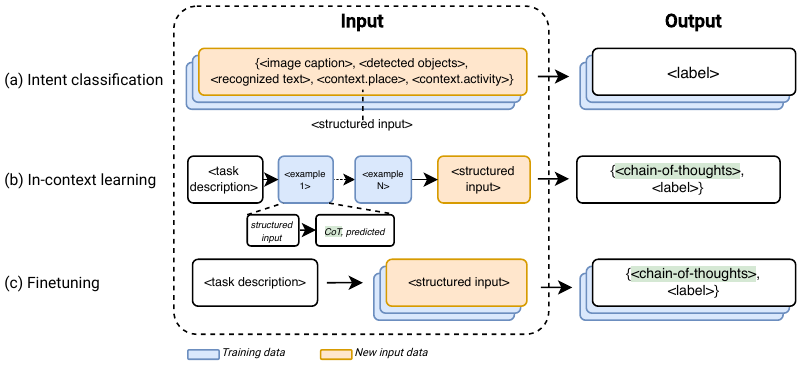}
    \caption{Data preparation and processing for each technique. \textit{Intent classification} and \textit{finetuning} used input-output pairs for training, while \textit{in-context learning} required only a few task examples.}
    \label{fig:techniques}
\end{figure*}

\subsubsection{In-Context Learning with CoT}

In-context learning, also known as few-shot prompting, is a popular method for adapting LLMs to new tasks \cite{brown2020language}.
This technique provides a few examples illustrating the task, specifying both the input format and expected output, without changing the model's parameters (\textit{i.e., }gradients) for new tasks. 
This is the key benefit that it does not require a large amount of data for training, thus making it potentially more adaptable to new tasks.

To enhance the explainability of the prediction, we provided exemplar data to instruct the LLM to produce intermediate reasoning (CoT) prior to the final action prediction.
We used both \texttt{GPT-3.5-turbo} and \texttt{GPT-4} as the model for the few-shot prompting method. 
As shown in Figure \ref{fig:techniques}, besides the converted structured text as the input, we also provide task descriptions and several examples illustrating the exemplar input and output. 
Specifically, the \textit{task description} defines the role of the system and leverages the definition of the predicted labels from the design space (\eg definition of specific actions in Figure \ref{fig:designspace}).
For the prediction of follow-up actions,
Since our design space consists of 17 specific categories, we include 9 data entries which cover all the categories in the prompt, and the rest 373 data entries are used for evaluation.
For detailed prompts, please refer to Appendix \ref{app:action}.

\subsubsection{Fine-Tuning with CoT}

Different from in-context learning, fine-tuning an LLM would change the model's parameters to specialize it for the target task.
This was accomplished by feeding additional training data into a pre-trained model, updating the model's gradients, \ie \textit{fine-tuning}.
The key benefit of this approach is that it enables the model to be exposed to a broader range of examples, and could thus potentially identify and learn more intricate patterns for better performance. 
However, the drawback is its reliance on a large amount of training data.

As shown in Figure \ref{fig:techniques}, for each data entry, we provided the structured text and the task description as the input and used the generated CoT and target label as the output. 
We used 75\% of the data entries for training and the rest for evaluation.
As GPT-4 did not publicly support fine-tuning at the time of this paper's preparation\footnote{as of December 11th, 2023}, we used \texttt{GPT-3.5-turbo} for the fine-tuning approach.

\end{colortext}

\begin{colortext}

\subsection{Performance Evaluation - Accuracy}


\label{sec:eval}


The two tasks: \one predicting the target information and \two predicting the follow-up actions, were performed in parallel and thus we evaluate them separately.

\subsubsection{Accuracy When Predicting Target Information}
Target information prediction is a \textit{multi-class classification} task, where
\end{colortext}the target modality was one of five modalities: 
the whole scene (\eg capture the whole moment or share a view with friends), physical objects (\eg recognizing a specific product and search online), the text visible in a visual (\eg save a promo code on a gift card), speech (\eg transcribe the teacher's lecture), or acoustic sound (\eg recognize background music).
As 80 diary entries were audio-only and there was only a text description of the audio without any visual information, we decided to separate the target modality prediction.
\begin{colortext}
Specifically, we implemented a \textit{three-class classification} (scenes, objects, and text) for visual information, and a \textit{two-class classification} (speech and sounds) for audio.\end{colortext}

\begin{table}[ht]
\centering
\caption{Accuracy (\%) when predicting the target information. 
}
\renewcommand{\arraystretch}{1.3} 
\begin{tabular}{ccc}
\hline

\centering{Approach} & Visual  & Audio \\ \hline

Intent classification          & 70.6            &     \textbf{92.3}       \\ 
In-context learning (w/ CoT)          & 62.3        &       90.1         \\ 
Fine-tuning (w/ CoT)                    &  \textbf{70.7}         &         90.9        \\

\bottomrule
\end{tabular}
\label{tab:target_accuracy}
\end{table}

\begin{colortext}

We measured the accuracy of the the three techniques. For \textit{intent classification} and \textit{finetuning}, we used 75\% of the data entries for training and the remaining 25\% for testing.
For \textit{in-context learning}, we used five data entries from the training set representing each target information modality as the few-shot examples and tested on the rest data (377 entries).
The results showed that all the approaches could achieve competitive performance when predicting the target information (Table \ref{tab:target_accuracy}). 

\end{colortext}




\begin{colortext}

\subsubsection{Accuracy When Predicting Follow-Up Actions}

\textcolor{black}{
As users may perform multiple actions using the same information, the prediction of follow-up actions is a \textit{multi-label classification} task, meaning each data entry may contain multiple ground truth labels.
Thus, we evaluated the model's accuracy when predicting the top-N most likely predictions (N = 1, 2, 3).}
It is important to note that, in the current setup, the accuracy of predicting the follow-up actions is not affected by the target information prediction as these two evaluations are conducted in parallel.
\textcolor{black}{
We used the \textit{full-match} metric to represent the accuracy of the prediction (i.e., the ratio of correct predictions to the minimum of ground truth labels or predictions), to demonstrate the alignment between the predictions and ground truth labels.} The accuracy was calculated using a sample average:

\begin{equation}
\text{Accuracy} = \frac{1}{N} \sum_{i=1}^{N} \frac{C_i}{\min(G_i, P_i)}
\end{equation}

where $N$ was the total number of test data samples, $C_i$ represented the number of correct predictions for the $i$-th data sample, $G_i$ represented the number of ground truth labels for the $i$-th data sample, and $P_i$ represented the number of predictions made for the $i$-th data sample.

\input{tables/comparison}

Besides the three approaches, we also calculated the accuracy of a model when it always predicted the top-N most frequent actions as it might achieve high accuracy due to imbalanced distribution of the data.
However, this does not make such a model \textit{good}, as it will never be able to predict actions other than the most dominant ones. Please refer to Appendix Figure \ref{fig:confusion_matrix_dominant} for the confusion matrix of this approach.

\paragraph{Results}
The results shown in Table~\ref{tab:accuracy_comparison} demonstrated that in-context learning with the latest LLM (GPT-4) outperformed all other approaches.
Notably, it achieves very high accuracy on general actions when predicting the top three possibilities (94.3\%) and 
marked an improvement of 11.6\% on specific actions over the next best-performant approach: fine-tuning with GPT-3.5 (from 60.1\% to 67.1\%).
Additionally, the results show that finetuning works better on specific actions (\textcolor{black}{13.8\%} improvement) than on general actions (\textcolor{black}{6.3}\% improvement) when predicting top-3 likely actions using the same model (GPT-3.5).
This is likely due to the dominance of certain categories in general actions and data-driven approach like finetuning is more sensitive to the data distribution. For detailed data, please refer to Appendix Table \ref{tab:improvement}.
\begin{figure*}[t]
    \centering
    \includegraphics[width=\linewidth]{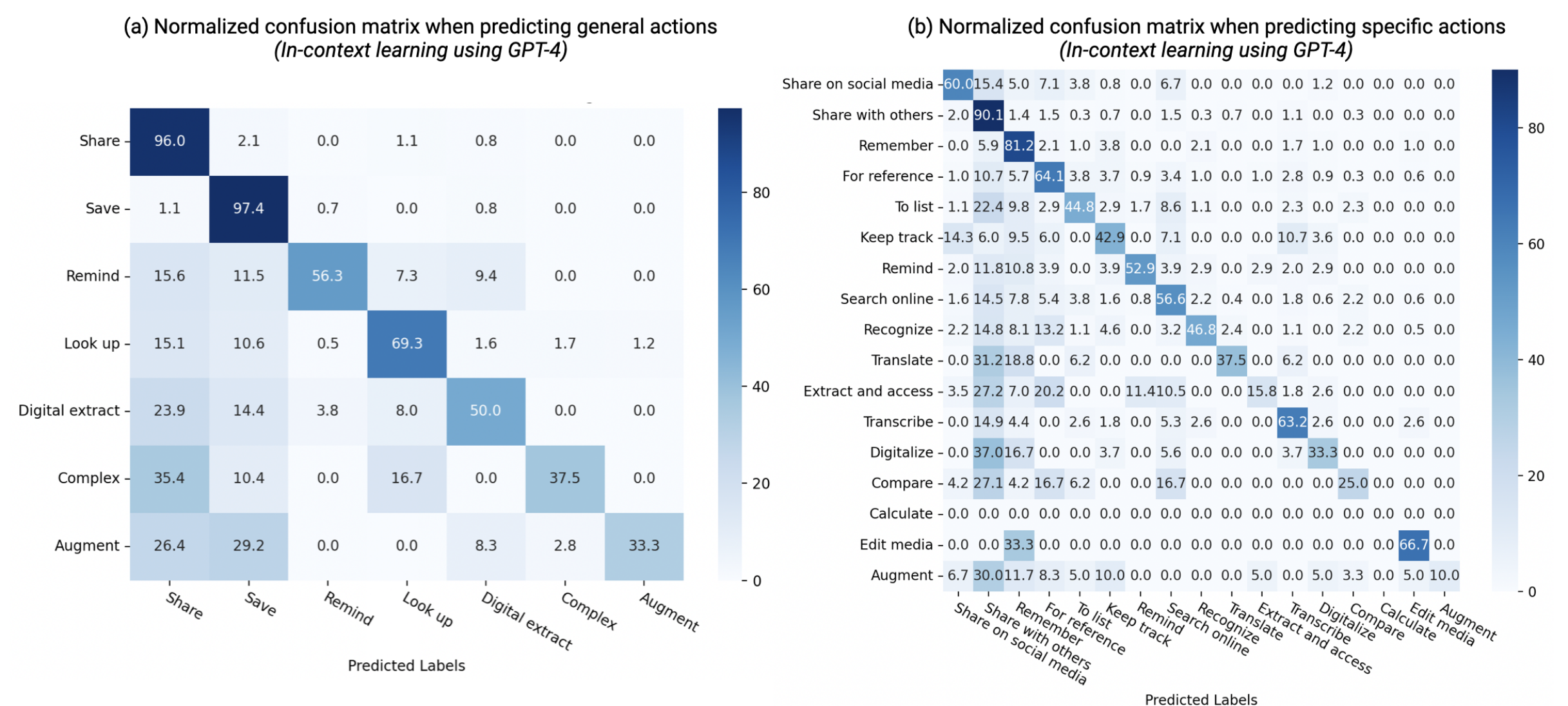}
    \caption{\begin{colortext}Confusion matrices using in-context learning (\texttt{GPT-4}) to predict the top-3 actions.\end{colortext}}
    \label{fig:confusion_matrices}
\end{figure*}

\subsection{Confusion Matrices of Predicting Follow-up Actions}

Besides the overall prediction accuracy, it is also important to analyze the error -- how does the model behave when predicting an incorrect label. 
We generated confusion matrices for the approaches to visualize the model behavior when predicting the top-3 actions (Figure \ref{fig:confusion_matrices}).
Specifically, we visualized the confusion matrices of the best-performant approach (\textit{i.e., in-context learning using GPT-4}) in this section.
Due to the imbalanced distribution of the data, we normalized the confusion matrix by the number of appearances of each label.
For details on creating these matrices and matrices for other approaches, please refer to Appendix \ref{app:confusion_matrix}.
Note that since we only have one data entry for the \textit{Calculate} action in the specific category and we have included that in the prompt, there is no data entry for this action in the evaluation set in this approach. 

\paragraph{Results}
The result shows a competitive performance using the in-context learning approach when sufficient examples are provided to cover the diversity of the actions.
This highlights the importance of \textit{data diversity} and the potentials for expanding the action space as interaction platforms and techniques evolve.
Regarding the data distribution, even without explicit awareness of it,  the model 
performs better on the dominant ones (\eg actions in the general \textit{share} and \textit{save} categories), while it performs worse on the less dominant ones (\eg specific actions like \textit{extract and access} or \textit{compare}). 
This shows an alignment between the data collected from the general users and the world knowledge that the model was trained on.
To increase the model's performance on less dominant categories, soliciting more data for certain actions might be necessary.
A future direction could involve collecting more high-quality data, which can be used to enrich the prompts for the in-context learning approach or employed for finetuning the model.

\begin{table}[ht]
    \centering
    \caption{Accuracy (\%) for the in-context learning approach with and without explicit contextual information while predicting three specific actions.}
    \renewcommand{\arraystretch}{1.3} 
    \begin{tabularx}{\columnwidth}{l>{\centering\arraybackslash}m{1.3cm}>{\centering\arraybackslash}m{1.3cm}>{\centering\arraybackslash}m{1.3cm}>{\centering\arraybackslash}m{1.3cm}}
    \hline
    & \textbf{W/O Context} & \textbf{Location Only} & \textbf{Activity Only} & \textbf{Full\newline Context} \\ \hline
    
    Audio only         & 47.5 & 47.7 & 59.7 & \textbf{60.0} \\ 
    Visual only        & 55.1 & 59.1 & 67.5 & \textbf{70.8} \\ \hline
    \textbf{All data}  & 52.5 & 55.2 & 64.9 & \textbf{67.1} \\
    \hline
    \end{tabularx}
    \label{tab:context_accuracy}
\end{table}

\subsection{Ablation to Understand Explicit Contextual Information and Modalities}
\label{sec:ablation}

The role of contextual information in the model's performance was another crucial aspect to consider. 
In our evaluation, we utilized data from the diary study assuming that the context was known, however, contextual information might not be readily available in practical scenarios. 
To understand its impact, we conducted an ablation test using the best-performing approach (\ie in-context learning with \texttt{GPT-4}), focusing on the two types of contextual information considered. We then computed the accuracy to assess the impact (Table \ref{tab:context_accuracy}).
Furthermore, we also examined how the model performs on visual and audio data separately to gain insights whether contextual information are important for certain modalities.

The result shows that the model's performance was improved by 27.8\% when the full context was provided compared to when no context was provided.
Within the contextual information, the activity information contributed more to the model's performance than the location information (23.6\% improvement for activity and 5.1\% for location), especially for audio data (25.7\% improvement). 
Besides the contextual information, the result also shows that the model performs generally better on visual data than audio data (70.8\% vs. 60.0\%).
This might be due to the richer content inherent in visual data, which contains more implicit contextual information. 
Recent research has shown multimodal models' capabilities in answering questions about the context from visual information \cite{dai2023instructblip}, thus future work may leverage such multimodal models to extract explicit contextual information before a prediction task.




\end{colortext}

%% file: tables/comparison.tex
\begin{table*}[ht]
    \centering
    \caption{Overall accuracy (\%) when predicting follow-up actions using the full-match metrics.
    }
    
    \renewcommand{\arraystretch}{1.1} 
    \begin{tabular}{C{4.1cm}*{6}{C{1.2cm}}}

      & \multicolumn{3}{c}{\textbf{Predicting General}} & \multicolumn{3}{c}{\textbf{Predicting Specific}} \\
     \hline

     Approach / Num of Predictions & 1  & 2 & 3 & 1  & 2 & 3  \\ 
    \hline
        
    \hline
    \textit{Top-n dominant categories}   & 47.4  & 61.3  &  78.1  & 39.3 &  45.3  &  54.8  \\
    Intent classification          & 46.0             &    61.1  &  83.1 &  41.7 & 40.6 & 54.3 \\ 
    Finetuning (GPT-3.5)          &  57.7       &       67.2    & 84.9 &  \textbf{48.1} & 50.2 & 60.1     \\
    In-context learning (GPT-3.5)         & 57.9             &     65.2  & 78.6 &  36.4 & 40.1 & 46.3  \\ 
    In-context learning (GPT-4)         & \textbf{60.3}     &     \textbf{69.9}  & \textbf{94.3} &  44.4  & \textbf{52.9}  & \textbf{67.1} \\
    \bottomrule
    \multicolumn{5}{l}{\footnotesize *All approaches except \textit{intent classification} adopt chain-of-thoughts.} \\
    \multicolumn{7}{l}{\footnotesize *Top-3 general actions (in order): Save, Share, Look up. Top-3 specific actions: Share with others, Save for reference, Search online.}\\
    \multicolumn{5}{l}{\footnotesize *In-context learning (GPT-4) is tested on 373 data entries.}
    \end{tabular}
    \label{tab:accuracy_comparison}
\end{table*}

%% file: content/06_system.tex

\section{A Mobile Proof-of-Concept Prototype with \codename Service}
\label{sec:system}


To give an example about how \codename' pipeline serve applications, we developed an interactive prototype (\ie an Android app), which passes the multimodal input to \codename and then executes the predicted follow-up actions.

\begin{figure}[ht]
    \centering
    \includegraphics[width=\columnwidth]{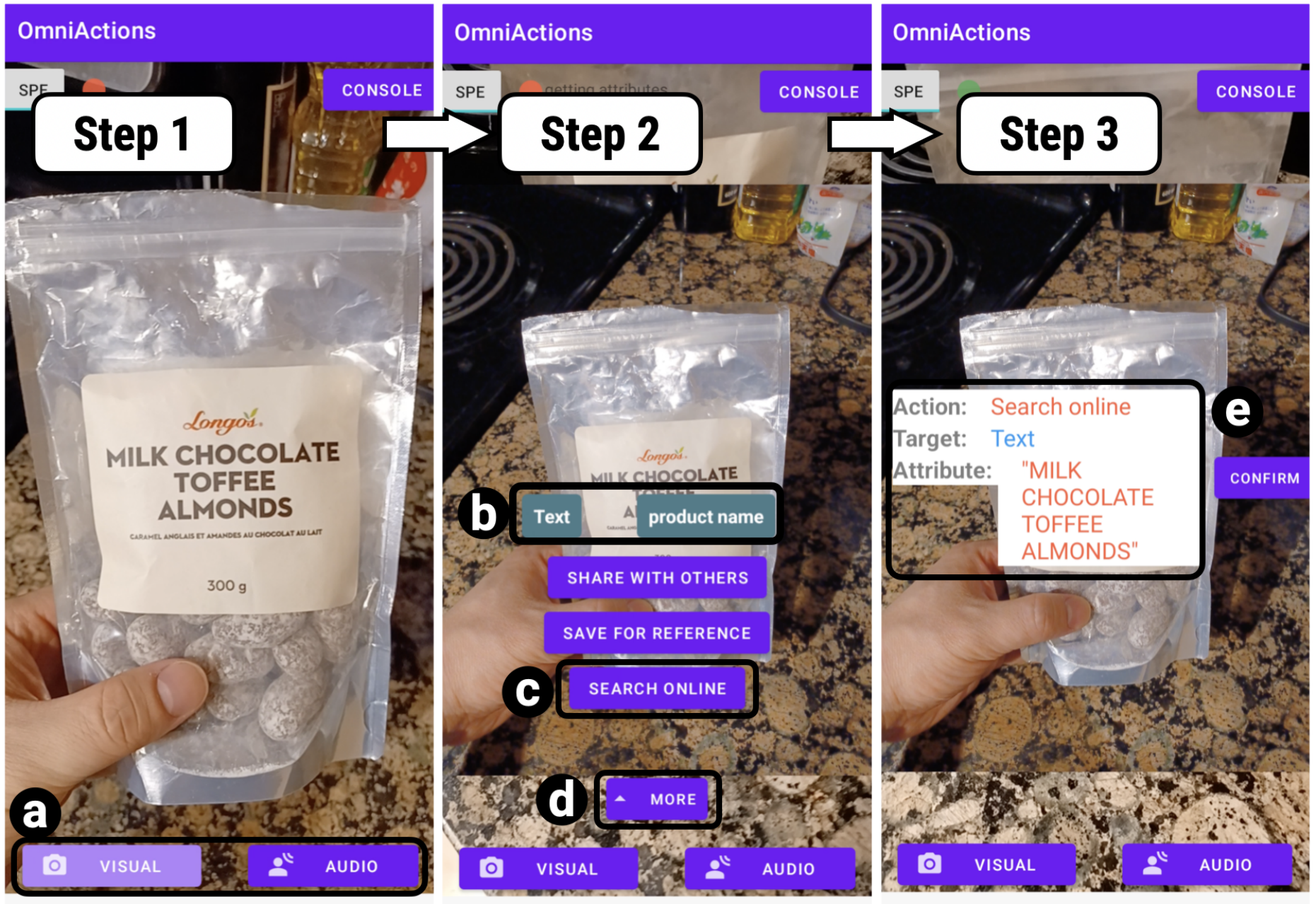}
        \caption{\textit{\codename}'s user interface, wherein (a-e) a user could search for the product name on the bag of chocolate by selecting the follow-up actions suggested by the system.
        }
    \label{fig:user_interface}
\end{figure}


\subsection{Workflow}

In this workflow, a user is searching for the product name of the chocolate online (Figure \ref{fig:user_interface}).
First, the user clicks the visual or audio button to specify the modality of information they are interested in.
As the user clicked the visual button (a), the system performs a \textit{target modality} prediction and \textit{follow-up action} prediction. The system then predicts the target as text (b) and recommends three actions.
If the user finds that the suggested actions do not fit their needs, they can click the \textit{more} button to see other actions in the design space (d).
The user then selects the target attribute of the text (``product name'') (b) and selects the \textit{Search Online} action (c).
 After selection, a pop-up window visualizes the user's intent to search for the product name (``MILK CHOCOLATE TOFFEE ALMONDS'') online (e).
 As the system does not currently detect all the context automatically, the user can manually specify a place and activity in the console (Figure \ref{fig:console}) for better prediction performance.
Additionally, the user can toggle between predicting general actions and specific actions to view the raw results to increase explainability in the console view as well. 

\subsection{Implementation}


The \codename prototype had two modules, a continuous detection module and a trigger-based detection module. The continuous detection module classified the sounds and transcribed speech (if present) in real-time and stored the classified sounds and speech transcription from the previous five seconds for further processing.
The trigger-based module captioned the captured images to provide a description, detected objects within the captured images, and used OCR to identify and extract text in the images.
Once a user triggered the system, \codename processed all the information into a tuple format so it could be used by the fine-tuned model for prediction.

The system was implemented on a Samsung Galaxy A13 5G phone running Android version 13.0. 
The code was developed in Android Studio using API level 33 and was written in the Kotlin programming language. 
The image captioning on the phone utilized the \texttt{blip-image-captioning-base} via the Hugging Face API, the object detection used \texttt{MobileNet V1}, and the text recognition used the Google Cloud Vision API. The audio classification used \texttt{YAMNet} and the continuous speech-to-text recognition used the Google Cloud Speech API.  




%% file: content/07_user_study.tex

\subsection{Preliminary User Feedback}
We used a think-aloud protocol \cite{nielsen1994usability} to understand how users perceive and use the prototype. Specifically, we are interested in people's reactions to the proactive interface and the prediction errors.  

\subsubsection{Setup and Method}
Five participants with either programming or product development experience were recruited from a our institution to participate in the study. The participants volunteered to join the study and they were not paid.
The study took place in a lab designed to resemble a cafe, which enabled everyday life scenarios such as viewing a menu and interacting with a book on a bookshelf.  
During the study, a researcher first walked the participants through the basic functionality by demonstrating an example.
Participants were then asked to complete six predefined tasks and verbalize their thoughts while doing so. 
\begin{colortext}
These include tasks such as ``\textit{save} the promocode on the gift card for future reference'' or ``\textit{share} the menu in a cafe with your friends''. 
\end{colortext}
Lastly, participants used the system to complete additional free-form tasks of users' choices (for as many times as they wanted), where they decided what follow-up actions they'd like to do. 
\begin{colortext}
Using the \textit{think-aloud} protocol, participants were asked to verbalize their intention on the actions they were taking and then used the system to complete the free-form tasks. 
\end{colortext}
After using the prototype system, participants completed a questionnaire containing 7 point Likert-based usability questions, as well as open-ended questions designed to gather qualitative feedback.  
the study took between 30-40 minutes to complete. We recorded audio during the study for later transcription and qualitative analysis.

\subsubsection{Results}
All participants successfully completed the predefined tasks without any assistance. 
Participants thought the system was easy to use (M = 4.8 ,$\sigma$=1.30), they were fond of it (M = 5.6, $\sigma$=1.34), and they thought it had potential and promise (M = 5.8, $\sigma$=1.64). 
As the participants experienced the proactive action prediction, they commented on how they could use \codename for their everyday tasks in the future. P3 stated, "\textit{having this might fundamentally change the interaction of future AR interfaces}". \codename was positively received due to its ability to reduce friction by predicting the actions (P1, P2, P4). 

Note that the system did not always predict the users' intended actions correctly. In these cases, the "more" function to quickly view other potential actions was used. For example, P3 commented that the \textit{"comprehensive overview of available actions was very useful."}. This showed the importance to have mechanisms to handle the scenarios where AI predictions didn't match users' intention. However, some users noted that visiting "more" actions could increase the cognitive load as there were many options to read and choose from. Participants (P1, P5) found it overwhelming to go through all the potential actions. To address this challenge, some participants suggested using hierarchical sub-menus (P1, P3, P5) or having fewer options while treating some actions as add-ons (P2). The hierarchical sub-menus could be supported by the prediction of the general actions (which had high accuracy) and then specific actions. 

Participants also shared areas of improvement for the prototype. One confusion area is the different interpretation of the wording for actions. For example, "\textit{I thought Save-to-list is saving something important to me while Save-for-reference is something that is not important}" (P3). P2 also stated "\textit{as a developer, I see the value of distinction between each actions which help me implement the functions ... but as an end-user, I find it confusing to differentiate between them and understand specific purposes}". P2 also mentioned that "\textit{trying to understand the difference between two suggested similar actions may also increase my cognitive load}". Participants suggested adding content-aware examples to each action to help end-users understand the outcome.
Overall, participants were enthusiastic about \codename, saw its value for end-users and developers, and provided suggestions for its improvement.

%% file: content/08_discussion.tex
\section{Discussion}
\label{sec:discussion}

In this section, we reflect on the design and evaluation of \codename. Our insights shed light on the design and implementation of proactive interfaces for AR use cases. We will also discuss the limitations of our current data and method, and a future direction to address these limitations.


\begin{colortext}

\subsection{Action Space for Everyday Information Encounters}
As far as we know, our work was the first to identify the set of actions people tend to take on the information they encounter during everyday tasks. The diary study method enabled us to capture moments of action needs in-situ, covering the majority of everyday activity types as context. In half of the cases, these activities involved high physical/social/cognitive effort, raising the importance to reduce the friction to any additional interactions. 
These everyday life scenarios captured in our dataset overlap with those in the Pervasive AR vision, where people use AR anytime and anywhere \cite{grubert2016towards}. Taking the lens of the Jobs-to-be-done \cite{christensen2016know}, each action was "hired" to address a human need, such as staying connected, getting emotional support, reducing memory load, and gaining more understanding, etc. While technology may be fast-evolving, human needs remain relatively stable. 

We understand, however, that actions shared by the participants in the current study are limited to actions they are familiar with on their current devices, specifically, phone-based actions. We expect these actions will be different when AR platforms are widely adopted and the ecosystems of actions on these platforms thrive. This kind of socio-technical co-evolution has been witnessed when we look into the literature about how people handled information needs with mobile phones decades ago. Back in 2008, people addressed information needs using the web, map, and calling on the phone, as well as through physical means (e.g. printing, asking something) \cite{sohn2008diary}. In contrast, our dataset shows a greater diversity of actions people can take than before, thanks to the fast evolution and wide adoption of smartphones. 
Given the accelerating pace of technology, we can also expect an increased number of capabilities and variety of actions supported by future AR platforms with an always-on sensor stack and increased computational intelligence. For example, the actions will be more adaptive to users' contexts with multi-sensor streams, more proactive with better prediction of users' intention from eye tracking, and more tailored to users' preferences and goals with the first-person perspective cameras etc. These new computing platforms will be able to provide users with different actions tailored to the individual to address their everyday needs in response to information triggers in the world. For the future systems that predict user's follow-up actions, developers will need to update their data over time to reflect the evolution of the actions, like any AI system would do.


\subsection{Actions Prediction with Multimodal Information}
With \codename, we created a pipeline that predicts follow-up actions and target information by turning the multi-modal information into structured texts to LLM. Among several state-of-art LLM techniques, we identified the most performant one (in-context learning with chain-of-thoughts, using enough examples to cover the diversity of the actions) that reaches competitive accuracy in prediction accuracy. 

Compared to multimodal LLMs (e.g. GPT-4v) where raw multimodal information was used directly as input, our approach has more transparency and explanability. We could evaluate how much each of the contextual factors contribute to end-to-end prediction performance by including/excluding it. We could better leverage chain-of-thoughts because the reasoning involves multiple contextual factors. We generate the chain-of-thoughts prompts from user data rather than a researcher's common sense. 
For all three LLM techniques we evaluated (intent classification, fine-tuning, and in-context learning), their performance relied on the dataset quality. Therefore, it is critical to collect up-to-date and relevant data that covers a wide range of the action space. As we mentioned in the above section, when the computing platforms like AR evolves, the action space will change, and the data needs to be updated. 


In our work, the collection of data from the diary study and the use of the data in prediction were two separate steps. We envision integrating the data with online training. Users wear a lifelogging system throughout the day (\eg RayBan Stories\footnote{https://www.ray-ban.com/usa}); it captures how people act upon what information over time and trains/prompts the model with the data. Users could also later reflect on the data, identify important information they missed, and label potential actions related to it. This way the pipeline will gain up-to-date and personalized data iteratively with the user.  

\subsection{Handling Predictions Errors}
Like many other AI-based predictions, our system makes errors. With our mobile prototype that leverages \codename to surface actions, we got valuable feedback about users' reactions and suggestions when the prediction did not match their intention. 
It is critical to have mechanisms to recover from error (the "Offer Simple Error Handling" rule \cite{shneiderman2005shneiderman}), however, we observed \textcolor{black}{in the user feedback sessions} that presenting a "more" button to list the rest of the actions \textcolor{black}{may increase} people's cognitive load. 
One way to reduce the cognitive load in error handling might be to leverage the higher-level grouping of actions, which achieved a high accuracy (94\%) in the general action prediction. 
This would then funnel users to the right categories of actions from which they could process a smaller set of sub-actions.

\end{colortext}

%% file: content/09_conclusion.tex
\section{Conclusion}


In this paper, we presented \codename, which predicts follow-up actions when users encounter multimodal information. 
To inform the design of \codename, we conducted a five-day diary study to understand of the design space of follow-up actions. 
Through the study, we identified 7 general categories of actions (i.e., \textit{share}, \textit{save}, \textit{remind}, \textit{look up}, \textit{digital extract}, \textit{media manipulation}, and \textit{complex actions}) and 17 specific follow-up categories of actions.

We then developed the \codename pipeline and prototype to predict follow-up actions for multimodal information powered by an LLM. The system harnessed the reasoning capabilities of LLMs by introducing intermediate reasoning steps (\ie CoT prompting). We evaluated three state-of-art LLM techniques, and the results indicated that integrating CoT prompting significantly improved the system's performance. Specifically, the model attained 94\% accuracy when predicting top three general actions when using in-context learning with CoT prompting. 
We then conducted a user study to understand users' feedback towards the action prediction and its errors. The findings demonstrated the potential of \codename and provided valuable insights into possible enhancements for systems alike.


%% file: content/99_appendix.tex
\newpage

\appendix

\setcounter{figure}{0}



\section{Prompt templates}

\subsection{Chain-of-Thoughts Prompts}
\label{app:cot}

\begin{quote}
    
\{``role'': ``system'', ``content'':

``You are an assistant that produces chain-of-thoughts analysis leading to reasons about why users take specific follow-up actions from a third-person perspective. You should operate under the assumption that the goal is not known to you.

Follow-up actions:
Share on social media: Share/upload on social platforms

Share with others: Send the info to specific entities

Remember: Cherish a specific experience/moment for later recall

For reference: Store information for later usage or consultation

To list: Add information to a designated, organized collection

Keep track: Record the development of a task or goal

Remind: Make an alert or notice to remember something later

Search online: Search for more information online related to specific goals

Recognize: Identify the information using specific tools (e.g., song names)

Translate: Translate text/speech from one language to another

Extract and access: Extract and utilize information from sources

Transcribe: Convert audio to text

Digitize: Transform information to a digital format for easier access

Compare: Compare similarity and difference between two sets of info

Calculate: Perform mathematical operations to solve a problem/task

Edit media: Enhance images or sounds to improve overall experience

Augment: Modify media files to accomplish a specific task

Output in a list of JSON dicts, where applicable:  "chain-of-thoughts", "prediction" (the follow-up actions)''
\}
\end{quote}

\subsection{In-Context Learning Prompts to Predict Target Information}
\label{app:target_info}

Predicting \textbf{visual} target information:
\begin{quote}
You are an assistant that predicts the target information that users take follow-up actions on when they encounter multimodal information using chain-of-thoughts analysis. 

The target information include three categories: scene, object, text:

scene: users would like to take actions on the whole visual content

object: users would like to take actions on specific physical objects they see

text: users would like to take actions on visible text in the scene
\\
Output the prediction result in a JSON dict, where applicable: "chain-of-thoughts", "prediction"
\end{quote}

Predicting \textbf{audio} target information:
\begin{quote}
    You are an assistant that predicts the target information that users take follow-up actions on when they encounter multimodal information using chain-of-thoughts analysis. 

    The target information include two categories: sound, speech:
    
    sound: users would like to take actions on acoustic sound they hear

    speech: users would like to take actions on someone's speech
    
    Output the prediction result in a JSON dict, where applicable: "chain-of-thoughts", "prediction"
\end{quote}

\subsection{In-Context Learning Prompts to Predict Follow-up Actions}
\label{app:action}

Predicting \textbf{specific} follow-up actions:
\begin{quote}
    
\{``role'': ``system'', ``content'':

``You are an assistant that predicts the follow-up actions users will take based on multimodal information input using chain-of-thoughts analysis. 
Provide up to\newline [NUM\_OF\_PREDICTION] most likely follow-up actions from the following options (with definition):

Follow-up actions:

[CATEGORIES]: [DEFINITION] (refer to Figure \ref{fig:designspace})


















Output in a list of JSON dicts, where applicable:  "chain-of-thoughts", "prediction" (the follow-up actions)''
\},

\{
``role'': ``user'',
``content'':
``<example 1>''
\}, 

\{
``role'': ``assistant'',
``content'':
``<result 1>''
\}, 

\{
``role'': ``user'',
``content'':
``<example 2>''
\}

\{
``role'': ``assistant'',
``content'':
``<result 2>''
\}
\end{quote}

Predicting \textbf{general} follow-up actions:
\begin{quote}
    
\{``role'': ``system'', ``content'':

``You are an assistant that predicts the follow-up actions users will take based on multimodal information input using chain-of-thoughts analysis. 
Provide up to [NUM\_OF\_PREDICTION] most likely follow-up actions from the following options (with definition):

(general) 

Share

(specific) 

Share on social media: Share/upload on social platforms

Share with others: Send the info to specific entities
\\

(general)

Save

(specific)

Remember: Cherish a specific experience/moment for later recall

For reference: Store information for later usage or consultation

To list: Add information to a designated, organized collection

Keep track: Record the development of a task or goal
\\

(general)

Remind

(specific)

Remind: Make an alert or notice to remember something later
\\

(general)

Look up

(specific)

Search online: Search for more information online related to specific goals

Recognize: Identify the information using specific tools (e.g., song names)

Translate: Translate text/speech from one language to another
\\

(general)

Digital extract

(specific)

Extract and access: Extract and utilize information from sources

Transcribe: Convert audio to text

Digitize: Transform information to a digital format for easier access
\\

(general)

Complex

(specific) 

Compare: Compare similarity and difference between two sets of info

Calculate: Perform mathematical operations to solve a problem/task
\\

(general)

Augment

(specific) 

Edit media: Enhance images or sounds to improve overall experience

Augment visual/audio: Modify media files to accomplish a specific task

Output the prediction result in a list of JSON dicts (the length will be the number of prediction), where applicable: "chain\_of\_thoughts", "prediction"

Output the general category''
\},

\{
``role'': ``user'',
``content'':
``<example 1>''
\}, 

\{
``role'': ``assistant'',
``content'':
``<result 1>''
\}, 

\{
``role'': ``user'',
``content'':
``<example 2>''
\}

\{
``role'': ``assistant'',
``content'':
``<result 2>''
\}
\end{quote}

\begin{colortext}

\section{Confusion Matrices for All approaches}
\label{app:confusion_matrix}
To compute the confusion matrices for each action category, for each data instance, we need to count both the corrected and incorrect predictions for the ground truth label. 
However, since we are forcing the model to predict the top-3 likely actions, this would introduce unavoidable \textit{errors} which do not reflect the model's performance.
To account for this, we only count the error when there exists at least one groud truth label that is not correctly predicted by the model.


The confusion matrices for the following approaches:
(1) only predicting top-3 dominant actions,
(2) intent classification,
(3) finetuning GPT-3.5,
(4) in-context learning with GPT-3.5
are shown in Appendix Figure \ref{fig:confusion_matrix_dominant} to \ref{fig:confusion_in_context}.

\begin{table}[ht]
    \centering
    \caption{
        Improvement (\%) on each action category from in-context learning to finetuning. 
    }
    \renewcommand{\arraystretch}{1} 
    \begin{tabular}{r|C{1.4cm}C{0.9cm}C{1.2cm}}
    \hline
    Predicting General Actions & In-context learning  & Finetuning  &  \textbf{Improvment}  \\ \hline
    
    \textbf{Share*}         & 82.7          &       96.7     &  \textbf{+16.9}  \\ 
    \textbf{Save*}         & 78.7             &     96.9    &  \textbf{+23.1} \\ 
    Remind     & 6.2             &     0    &  -100 \\ 
    \textbf{Look up*}   & 66.9             &     93.4    &  \textbf{+39.6} \\ 
    Digital Extract & 56.4             &    17.9    &  -68.2 \\ 
    Complex & 12.5             &     0    &  -100 \\ 
    Augment & 40.0             &     20.0    &  -50.0 \\  \hline

    Predicting Specific Actions \\ \hline
    Share on social media & 78.4             &     4.5    &  -94.3 \\ 
    \textbf{Share with others*}   & 44.9             &     89.5    &  \textbf{+99.3} \\ 
    Remember & 70.2             &     47.8    &  -31.9 \\ 
    \textbf{For reference*} & 26.8             &     74.2    &  \textbf{+176.9} \\ 
    \textbf{To list} & 19.2            &     58.8   &  \textbf{+206.2} \\ 
    Keep track & 28.6             &     11.1    &  -61.2 \\ 
    Remind & 6.2            &    0   &  -100 \\ 
    \textbf{Search online*} & 64.6             &     70   & \textbf{ +8.4} \\ 
    \textbf{Recognize} & 25.9             &    56.7    &  \textbf{+118.9} \\ 
    Translate & 37.5           &     25   &  -33.3 \\ 
    Extract and access &   11.1            &     8.3   &  -25.2 \\ 
    Transcribe &   61.1            &    16.7    &  -72.7 \\ 
    Digitalize & 16.7            &    0  &  -100 \\ 
    Compare &   14.3             &    0    &  -100 \\ 
    Calculate & 0             &     0    &  0 \\ 
    Edit media & 0             &     0   &  0 \\ 
    \textbf{Augment} &  10.0            &     33.3   &  \textbf{+233.0} \\ 
    \hline
    \multicolumn{4}{l}{\footnotesize \textbf{Bolded} denotes positive improved categories.}
    \end{tabular}
    \label{tab:improvement}
\end{table}

Table \ref{tab:improvement} shows the improvement from in-context learning to finetuning using the same model (GPT-3.5). The results indicate that the finetuning method is sensitive to the distribution of training data. 
Notably, in the case of general actions, the dominant categories are excessively predominant (>30\%) accounting compared to other categories (<15\%). Conversely, in specific actions, the data is more evenly spread across various non-dominant categories.
Consequently, given the current data distribution, finetuning demonstrates better performance with specific actions than with general actions.

\end{colortext}






\section{Generating The Design Space}
\label{app:formative}


\begin{figure*}
    \centering
    \begin{subfigure}[t]{\linewidth}
        \centering
        \includegraphics[width=0.7\textwidth]{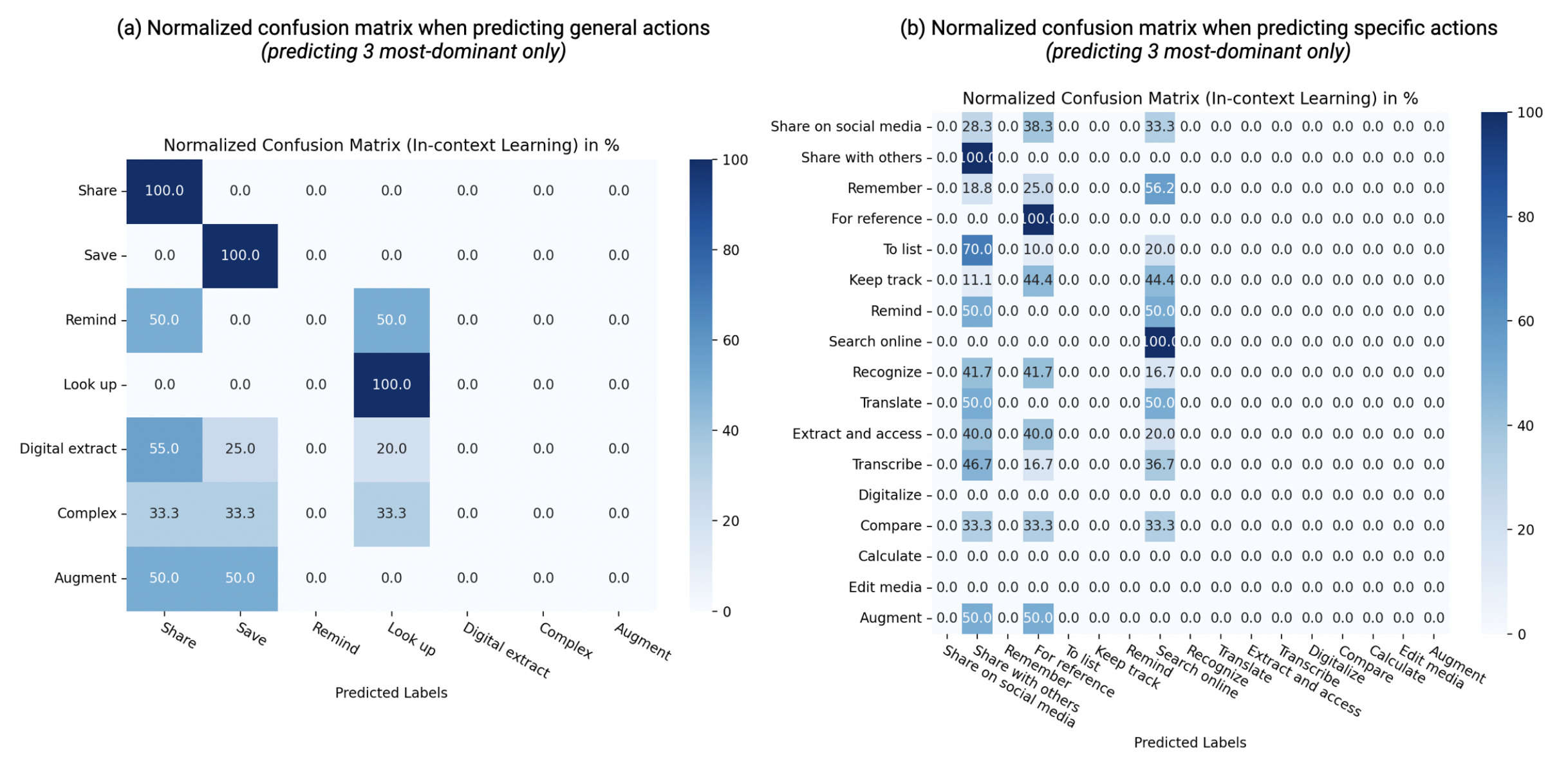}
        \caption{Confusion matrix for only predicting the dominant actions.}
        \label{fig:confusion_matrix_dominant}
    \end{subfigure}
    \begin{subfigure}[t]{\linewidth}
        \centering
        \includegraphics[width=0.7\textwidth]{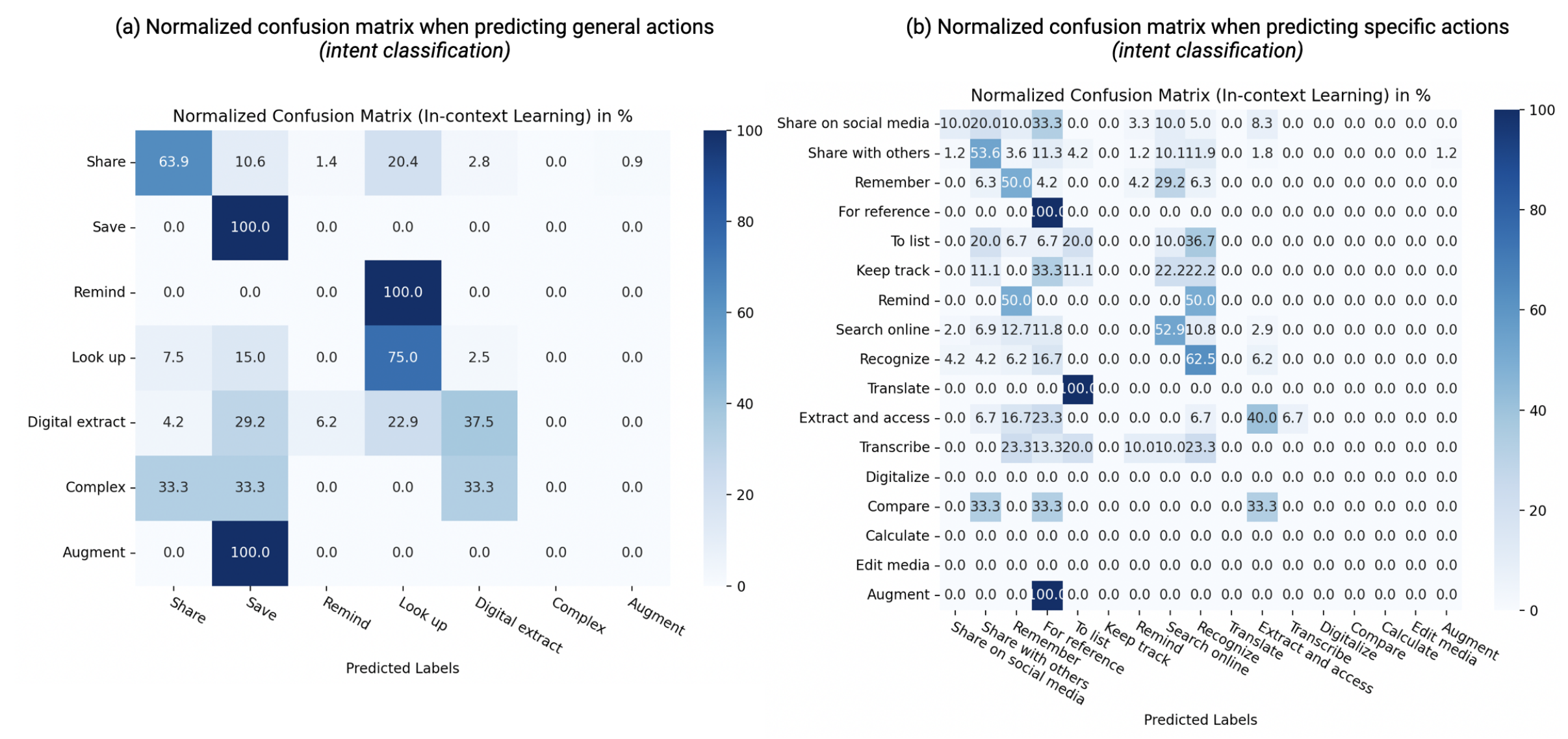}
        \caption{Confusion matrix for intent classification.}
        \label{fig:confusion_intent}
    \end{subfigure}
    
    \caption{Confusion matrices for predicting dominant only and intent classification.}
\end{figure*}

\label{app:diary_study}
\subsection{Survey Questions for the Diary Study}
The survey questions are listed in Table \ref{tab:survey_question}.

\begin{figure*}
    \centering
    \begin{subfigure}[t]{\linewidth}
        \centering
        \includegraphics[width=0.7\textwidth]{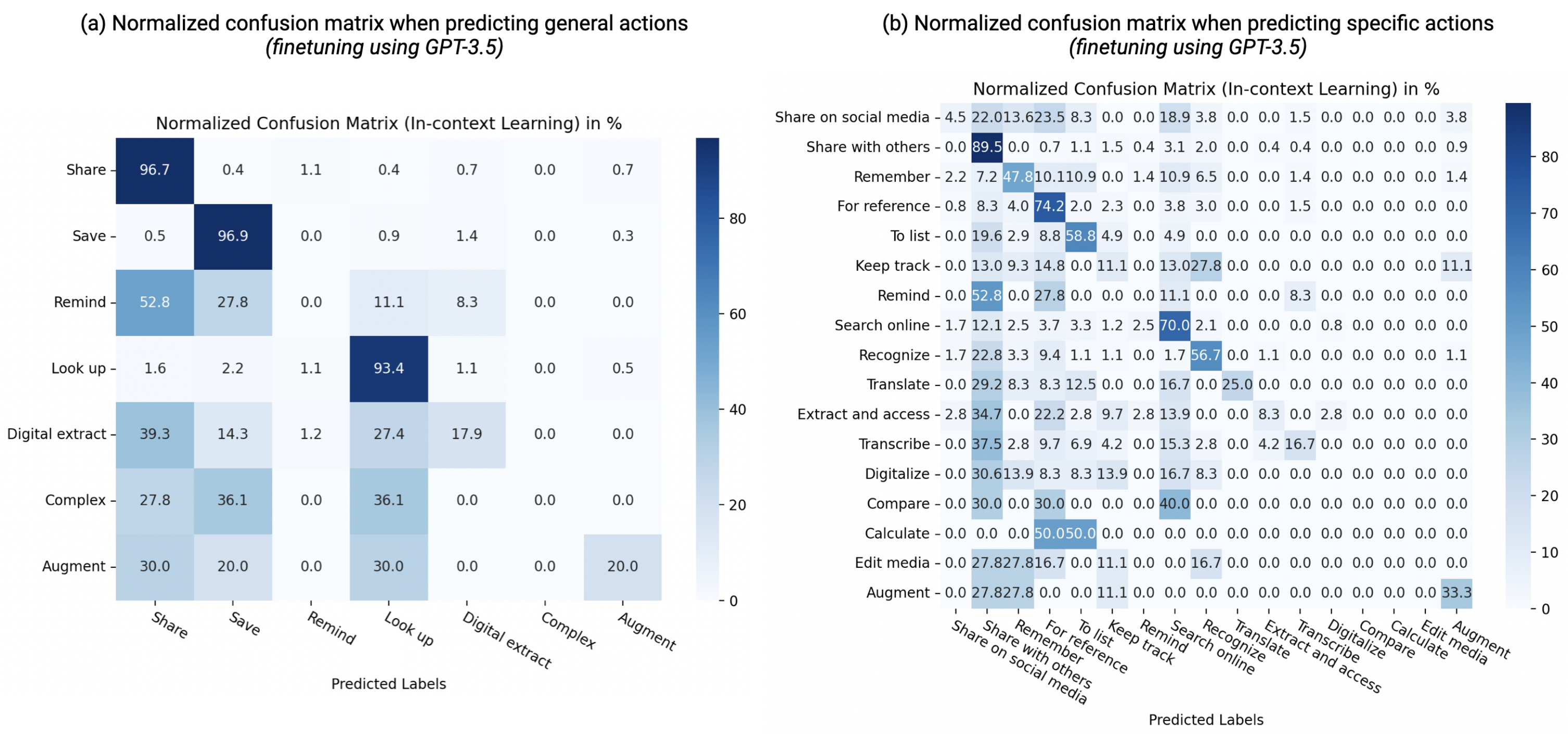}
        \caption{Confusion matrix for finetuning with GPT-3.5.}
        \label{fig:confusion_finetuning}
    \end{subfigure}
    
    \begin{subfigure}[t]{\linewidth}
        \centering
        \includegraphics[width=0.7\textwidth]{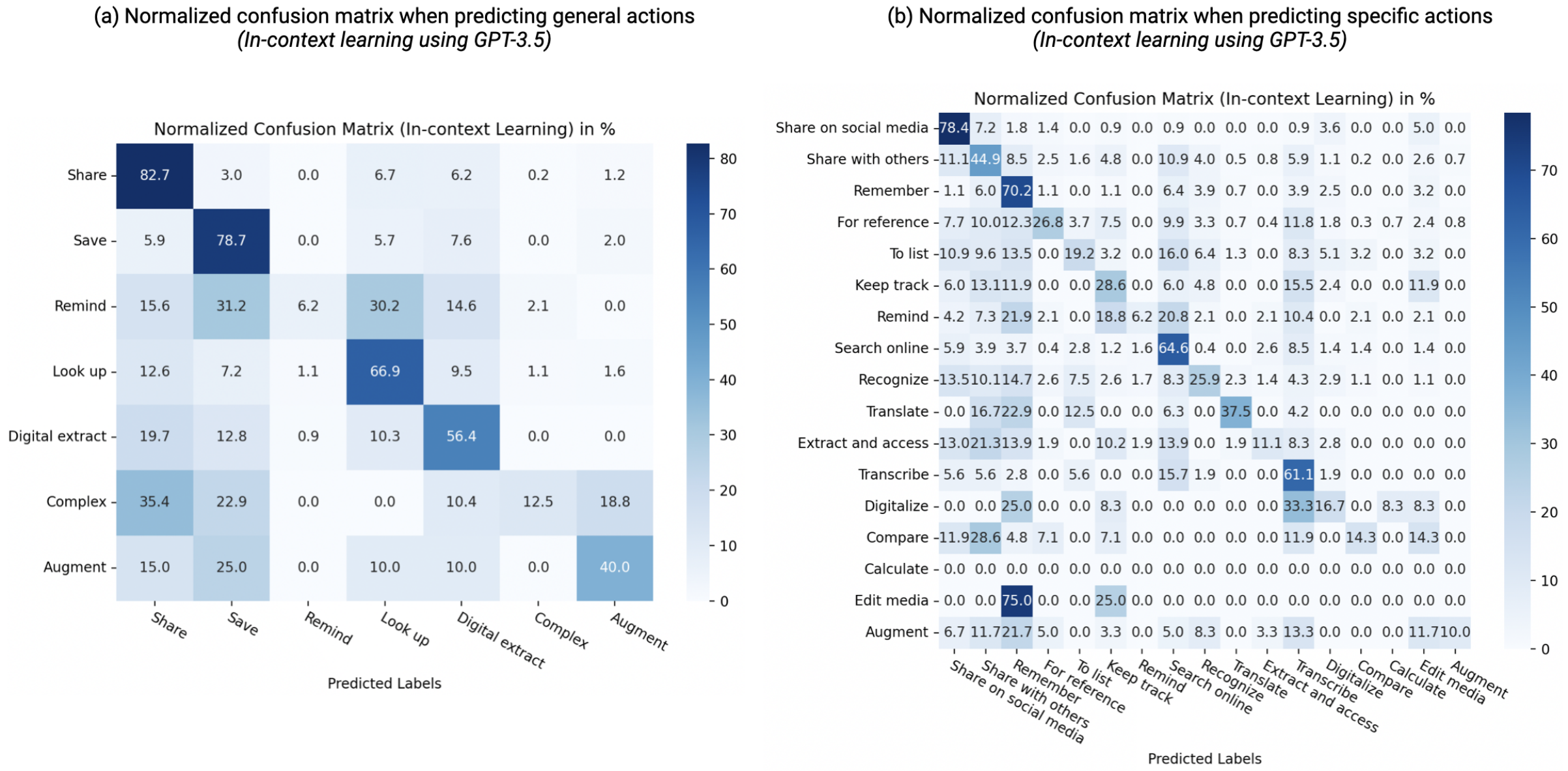}
        \caption{Confusion matrix for in context learning with GPT-3.5.}
        \label{fig:confusion_in_context}
    \end{subfigure}
    
    \caption{Confusion matrices for finetuning and in-context learning.}
\end{figure*}

\input{content/05_02_definition_of_actions}

\newpage
\section{Data with Aggregated Actions}
\label{app:aggregated}
\begin{figure}
    \centering
    \includegraphics[width=1\columnwidth]{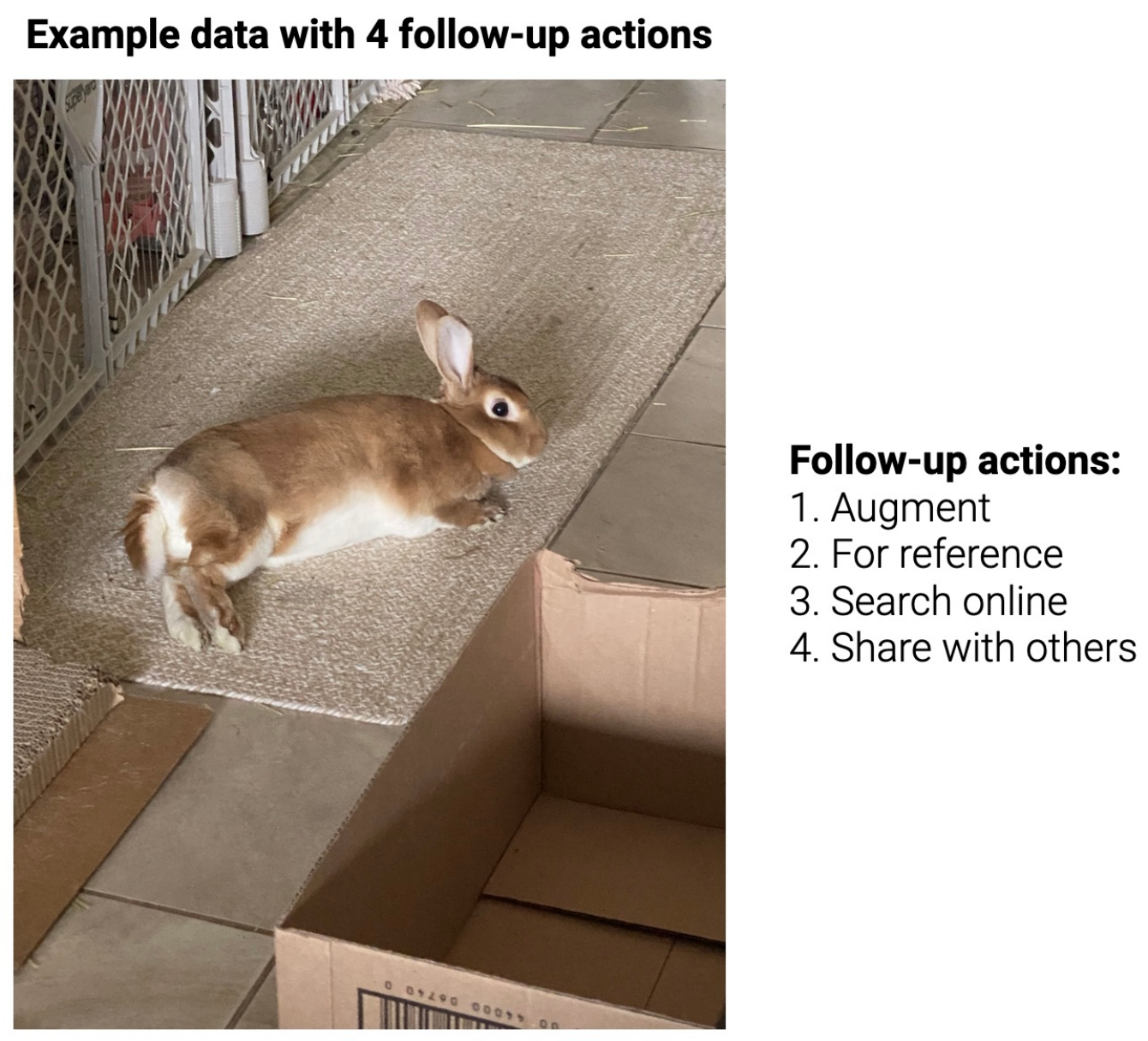}
    \caption{An example of the collected data with four follow-up actions.}
    \label{fig:aggregated_example}
\end{figure}

\begin{colortext}
Participants tends to perform multiple actions on the information they encounter. Figure \ref{fig:aggregated_example} shows an example of the collected data with four follow-up actions.
In this example, the participant took a picture of their rabbit as they think the rabbit might be ill.
Since the rabbit will run away if they get too close, the participant decided to take a picture of the rabbit first from afar to (1) zoom in for clearer view (\textit{augment}) and (2) share the picture with a veterinarian (\textit{share with others}).
They would also save the picture for future reference (\textit{for reference}) and could possibly search online for more information if the veterinarian is not available (\textit{search online}).

Table \ref{tab:aggregated_performance} shows performance of the model on data with and without aggregated actions.
\end{colortext}

\begin{table}[htp]
    \centering
    \caption{
        Accuracy (\%) on data with and without aggregated actions (predicting top-3 actions using in-context learning)
    }
    \renewcommand{\arraystretch}{1.3} 
    \begin{tabular}{ccccc|c|c}
    \hline
   Num of actions in data &  1 & 2  & 3 & 4 & >2 & All \\ \hline
    
    General Actions         & 98.7     &       91.2     &  68.6  & 87.5  & 85.5 & 94.3  \\ 
    Specific Actions         & 73.7     &    64.1    &  50.3 &  79.2 &  61.1 & 67.1 \\
    \hline
    \end{tabular}
    \label{tab:aggregated_performance}
\end{table}


\input{figures/diary_examples}

\input{tables/survey_questions}

%% file: content/05_02_definition_of_actions.tex
\subsection{Definition of Specific Follow-Up Action Categories}
\label{sec:category}
\subsubsection{\textbf{Share}}
%

\paragraph{Sharing with Others} 
When \textit{sharing with others}, future systems could leverage additional contextual information such as recommending people who have recently expressed their love for dogs when a user takes a photo of their dog.

\paragraph{Sharing on Social Media}
When \textit{sharing on social media}, future systems could suggest multiple hashtags to use. 

\subsubsection{\textbf{Save}}

\paragraph{Remember}
This refers to actions where users wished to cherish a specific moment to retrieve it in the future. \textit{Remember} often occurred when participants mentioned words such as ``funny'', ``memorable'', etc. or alongside other \textit{share} actions.

\paragraph{Save for Reference}
This refers to actions where users stored information with the specific goal of using it later. 
Participants mentioned various types of \textit{later usages}, including using it for a later purchase, saving a gift card to avoid losing it, and so on. 
By automatically incorporating metadata into the information (\eg when, where, and what type of object), future systems could enhance user experiences by enabling quick and efficient retrieval of the information when needed.

\paragraph{Save to a List}
These actions added information to a designated collection, e.g., music to a playlist. Future systems could leverage this action by identifying the category of the information (\eg painting, music, groceries, etc.) and store the information in a list.

\paragraph{Keeping Track of Progress}
Participants captured information to record their performance or progress towards specific goals such as recording the progress of their bulking (or cutting) while working out or playing the piano. 
Different from \textit{saving to a list}, this information tended to be similar yet sequential in nature, enabling users to observe and evaluate their growth over time, which could be supported by future systems.

\subsubsection{\textbf{Look Up}}

\paragraph{Search Online}
Users conducted online searches to acquire additional information related to their intent, utilizing a variety of search tools (\eg Google). 

\paragraph{Recognize}
Users also identified information using specific tools, e.g., product searching (\eg using Google Lens or Images) or recognizing music (\eg Shazam). 

\paragraph{Translate}
In the context of text or speech, \textit{translate} refers to the actions that sought the meaning of text or speech in a different language, enabling one to better understand and communicate across language barriers.

\subsubsection{\textbf{Digital Extract}}

\paragraph{Extract and Access}
These actions extracted information from the physical world and directly took action on it based on its type. For example, systems could enable users to directly scan and access the content of a QR code, take a picture of a contact card and directly make a phone call, or extract an address from text and navigate to it. 

\paragraph{Transcribe}
Mostly applying to audio, \textit{transcribe} refers to actions that converted audio into text. This included transcribing a lecture or transcribing the lyrics from a song that was playing.

\paragraph{Digitize} 
These actions transformed various forms of information, such as physical documents or audio, into a digital format for easier access, storage, or sharing. The most common \textit{digitize} actions scanned physical information to create a digital copy for easier access and sharing. Digitizing audio, for instance, involved converting voice recordings into digital files, which could then be added to various media, such as TikTok videos.

\subsubsection{\textbf{Media Manipulation}}

\paragraph{Augment Media}
\textit{Augment} refers to actions that enhanced images or sounds to improve overall experiences. For example, participants wanted to zoom in to see the details of an object or isolate music from noise for precise recognition. 

\paragraph{Edit Media}
This refers to actions that were taken to modify media files for specific tasks. For example, a participant wanted to trim a video to share it on social media. Another participant wanted to crop an image for her slides. 
These editing actions ranged from simple adjustments, such as cropping or resizing, to more complex alterations, such as color grading or adding visual effects. 

\subsubsection{\textbf{Complex Actions}}

\paragraph{Compare}
\textit{Compare} refers to actions that compared similarities and differences between two sets of information. One participant, for example, wanted to compare the price of two similar products. This would require a system to retrieve additional information and present it simultaneously for the user to compare. 

\paragraph{Calculate}
While only mentioned by one participant, \textit{calculate} actions involved performing mathematical operations to solve a problem or a task, e.g., calculating if the calories one consumed exceeded their daily limit while cutting weight.

%% file: figures/diary_examples.tex

  





%% file: tables/survey_questions.tex
\begin{table*}[t]
\centering
\caption{Survey questions that participants were required to answer for each diary entry.}
\label{tab:survey_question}
\begin{tabularx}{\textwidth}{c|X | X|c}
\hline
 \# & \centering Target: \textbf{Visual} & \centering Target: \textbf{Audio} & Question type \\
\hline
Q1 & Upload your photo or a screenshot of your video. & 
(For video only) Upload a screenshot of your video (audio as the main target). & [File upload] \\
\cline{2-3}
Q2 & 
Briefly describe the photo. \newline
\textit{\eg “This is a billboard of the movie Dunkirk showing when it will be in theater.”}
& 
Briefly describe the audio you captured AND wanted to take follow-up actions with. \newline
\textit{\eg ``This is the background music I heard in the cafe.''}
& [Open-ended] \\
\cline{2-3}
Q3 & 
\multicolumn{2}{l|}{Where were you when you captured the data?} 
& [Open-ended] \\
\cline{2-3}
Q4 & \multicolumn{2}{l|}{What were you doing when you captured the data?} & [Open-ended] \\
\cline{2-3}
Q5 & Please list the physical objects visible in the data. & What types of sounds could be heard in the recording? \newline
\textit{- Speech / Music / Tools / Environmental noise / ...} \newline
\textit{- Others [Force answer]} & [Multi-type]

\\
\cline{2-3}
Q6 & What best describes the information you intended to take action on? \newline 
\textit{- The whole scene / environment / place} \newline
\textit{- Objects in the photo/video} \newline
\textit{- Text visible in the photo/video} \newline
\textit{- Others [Force answer]}
&
Please choose the audio information you want to take action on: \newline
\textit{- [Same as in Q5]}
& [Multiple choice]\\

\cline{2-3}
Q7 & \multicolumn{2}{>{\hsize=\dimexpr2\hsize+2\tabcolsep+\arrayrulewidth\relax}X|}{In 1-3 sentences, explain what actions you plan to take on the information in the data you shared. \newline
\textit{For example: “Save the date to my calendar.” If you have multiple actions, please list them all.} } & [Open-ended] \\

\cline{2-3}
Q8 & \multicolumn{2}{>{\hsize=\dimexpr2\hsize+2\tabcolsep+\arrayrulewidth\relax}X|}{From the list below, which best characterizes your previous response. Select all that apply. \newline
\textit{- [Categories from the workshop]} \newline
\textit{- Others [Force answer]}
} & [Multiple choice] \\

\cline{2-3}
Q9 & \multicolumn{2}{>{\hsize=\dimexpr2\hsize+2\tabcolsep+\arrayrulewidth\relax}X|}{


In 1-3 sentences, briefly explain: \newline
\one the overall goal(s) of taking the above actions. \newline
\two the reason(s) why you want to take the above actions when you captured the photo/video.} & [Open-ended] \\


\hline
\end{tabularx}
\end{table*}